\documentclass[a4paper,12pt]{article}
\pdfoutput=1
\usepackage{graphicx, rotating,amssymb}

\long\def\symbolfootnote[#1]#2{\begingroup\def\thefootnote{\fnsymbol{footnote}}\footnote[#1]{#2}\endgroup}

\ifx\pdfoutput\undefined
\usepackage[dvips,bookmarks]{hyperref}	
\else
\usepackage{hyperref}	
\fi
\hypersetup{colorlinks,bookmarksopen,bookmarksnumbered,citecolor=rossoc,
linkcolor=verdes,pdfstartview=FitH,urlcolor=rosso}
\def\myurl#1#2{\href{http://#1}{#2}}
\def\hhref#1{\href{http://arxiv.org/abs/#1}{#1}} 
\def\mhref#1{\href{mailto:#1}{#1}}		

\usepackage{multicol}
\usepackage{color}
\definecolor{rosso}{cmyk}{0,1,1,0.4}
\definecolor{rossos}{cmyk}{0,1,1,0.55}
\definecolor{rossoc}{cmyk}{0,1,1,0.2}
\definecolor{blu}{cmyk}{1,1,0,0.3}
\definecolor{blus}{cmyk}{1,1,0,0.6}
\definecolor{bluc}{cmyk}{1,1,0,0.1}
\definecolor{verde}{cmyk}{0.92,0,0.59,0.25}
\definecolor{verdec}{cmyk}{0.92,0,0.59,0.15}
\definecolor{verdes}{cmyk}{0.92,0,0.59,0.4}

\newcommand{\beq}{\begin{equation}}
\newcommand{\eeq}{\end{equation}}

\newcommand{\GeV}{\,{\rm GeV}}
\newcommand{\TeV}{\,{\rm TeV}}

\newcommand{\DM}{{\rm DM}}

\def\Fermi{{\sc Fermi}} 
\def\PAMELA{{\sc Pamela}}
\def\AMS{{\sc Ams-02}}
\def\HESS{{\sc Hess}}

\begin{document}
\begin{flushright}
\footnotesize
SACLAY--T13/006
\end{flushright}
\color{black}
\vspace{0.3cm}

\begin{center}
{\LARGE\bf Antiprotons from Dark Matter:\\[2mm] current constraints and future sensitivities}

\medskip
\bigskip\color{black}\vspace{0.6cm}

{
{\large\bf Marco Cirelli}\ $^{a}\, \symbolfootnote[2]{\mhref{marco.cirelli@cea.fr}}$,
{\large\bf Ga\"elle Giesen}\ $^{a}\, \symbolfootnote[3]{\mhref{gaelle.giesen@cea.fr}}$
}
\\[7mm]
{\it $^a$ \href{http://ipht.cea.fr/en/index.php}{Institut de Physique Th\'eorique}, CNRS, URA 2306 \& CEA/Saclay,\\ 
	F-91191 Gif-sur-Yvette, France}\\[3mm]
\end{center}

\bigskip

\centerline{\large\bf Abstract}
\begin{quote}
\color{black}\large
We systematically analyze the impact of current and foreseen cosmic ray antiproton measurements on the properties of Dark Matter (DM). We find that: 1) The current data from \PAMELA\ impose constraints on annihilating and decaying DM which are similar to (or even slightly stronger than) the most stringent bounds coming from \Fermi\ gamma rays, for hadronic channels and with fiducial choices for the astrophysical parameters.   2) The upcoming data from \AMS\ have the power to improve these constraints by slightly less than one order of magnitude and even to probe the thermal relic DM in the range 30$-$200 GeV, for hadronic channels. 
However, with wider choices for the astrophysical parameters the uncertainty on the constraints spans between one and two orders of magnitude.
We then explore the capabilities of early \AMS\, data to reconstruct the underlying DM properties in the case of a positive detection of a significant excess (attributed to DM annilations) over the background. For hadronic channels, we find that \AMS\ should be able to somewhat determine the DM mass and the cross-section, but not the specific annihilation channel nor the branching ratios. If other more exotic annihilation channels are allowed, the reconstruction will be more challenging.
\end{quote}


\section{Introduction}
Overwhelming evidence for the existence of Dark Matter (DM), in the form of an unknown particle filling the galactic halos, originates from many observations in astrophysics and cosmology: its {\em gravitational} effects are apparent on galactic rotations, in galaxy clusters and in shaping the large scale structure of the Universe~\cite{BertoneReview}. On the other hand, a {\em non-gravitational} manifestation of its presence is yet to be unveiled. One of the most promising techniques is the one of {\em indirect detection}, aimed at identifying excesses in cosmic ray fluxes ($e^\pm, \gamma, \nu, \bar p, \bar d$) which could possibly be produced by DM annihilations or decays in the Milky Way halo. The current experimental efforts mainly focus in the GeV to TeV energy range, which is also where signals from WIMPs (Weakly Interacting Massive Particles) are expected.

Concerning $e^\pm$, the well known excesses detected by \PAMELA, \Fermi\ and \HESS\ have been indeed interpreted in terms of DM (see e.g.~\cite{CKRS}, among many) but then, given the tensions or plain inconsistencies with gamma ray and other constraints (see e.g.~\cite{Cirelli:2012tf} for a concise review), a more mundane explanation in terms of some previously-unaccounted-for astrophysical process now seems preferred. In turn, this means that $e^\pm$ lose their appeal as far as DM searches are concerned: the large excesses likely drown the possible signal from all but the most extreme (very large cross section, very large mass) DM models. 

In $\gamma$-rays searches, with a few exceptions, only constraints on DM are reported, from the observation of many different targets and in different bandwidths (see e.g.~\cite{Bringmann:2012ez} for an up-to-date overview). One notable exception consists of the claim of detection of a $\gamma$-ray line at around 135 GeV from the Galactic Center (and possibly elsewhere) in \Fermi\ data. This feature, however, needs further scrutiny and we will not further address it here. 

In neutrino searches, only constraints are reported, both from observation of the Galactic Center or the Sun (see again~\cite{Cirelli:2012tf} for references).

\medskip

Within this thriving area of research, we will focus on antiproton searches, for several reasons. First, antiproton production is relatively universal in DM models: as long as DM annihilates (or decays) into quarks or gauge bosons, $\bar p$ will be produced abundantly in the hadronization process, and even in leptophilic channels some $\bar p$ are generated, as we will discuss below. On the other hand, the yield of $\bar p$ from galactic astrophysical processes is relatively less important, a fact which makes for a potentially very good signal/background ratio.
Second, the prediction of $\bar p$ flux from astrophysical processes suffers from a relatively small uncertainty (as we will recall later) and the error band is particularly narrow around tens of GeV, which is precisely a fortunate window for the emergence of a DM signal. Third, the propagation of $\bar p$ in the Galaxy is relatively under control and is generically less dependent on the details of the galactic geography then, say, $e^\pm$ propagation. Finally, data on $\bar p$, currently from \PAMELA, are exquisitely precise and even better is foreseen for \AMS, so that the constraining power can be predicted to be very good.

\medskip

The aim of this paper is therefore to investigate the constraints and the projected sensitivity to DM signals in the $\bar p$ channel. 
A selection of past works with a similar aim, usually in some specific model or with the \PAMELA/\Fermi/\HESS\ $e^\pm$ anomalies in mind, includes~\cite{Ibarra:2008qg,CKRS,Donato:2008jk,Bringmann:2009ca,Garny:2011cj,Cerdeno:2011tf,Kappl:2011jw,Garny:2011ii,Asano:2011ik,Chu:2012qy,Chun:2012yt}. A particularly thorough analysis has been done in~\cite{Evoli:2011id} (with which we will compare later on). 
In this work we aim at a more systematical and model-independent analysis of antiproton capabilities than most previous works.
A first step in this direction has been made in~\cite{Belanger:2012ta}: we here extend to several different channels and to a large range of DM masses. 

\medskip

The rest of this paper is organized as follows. In Sec.~\ref{method} we recall the basics of the production and propagation of antiprotons (from DM as well as from astrophysics) and the choices we make for the analysis. In Sec.~\ref{results} we present our results: we first compute the current constraints from \PAMELA\ data; we then derive the sensitivities expected from \AMS\ data (if no signal will be seen) and finally we investigate the reconstruction capabilities of \AMS\ itself assuming that a positive signal will be seen. Finally in Sec.~\ref{conclusions} we conclude.

\section{Antiprotons from DM and astrophysics}
\label{method}

In this Section we briefly recall the basics of production of antiprotons in DM annihilations or decays (`primary antiprotons') and their propagation in the galactic halo of the Milky Way. We also specify our assumptions for the astrophysical $\bar p$ background (`secondary antiprotons'). 

\subsection{Primaries}

The term `primary antiprotons' refers to the $\bar p$ flux produced by DM annihilations or decays in the galactic halo. While we here briefly review the basic points focussing in particular on the choices we adopt, we refer for any other detail to the standard literature on the subject (e.g.~\cite{PPPC4DMID} and references therein). 

We consider several different annihilation or decay channels, in a model independent way. We start by considering seven of them: 
\beq
\left.
\begin{array}{rr}
{\rm annihilation} & {\rm DM} \ {\rm DM}  \\
{\rm decay} & {\rm DM}
 \end{array}\right\}
\to \ b \bar b, t \bar t,  W^+W^-,  ZZ, \mu^+ \mu^-, \tau^+ \tau^-, \gamma \gamma .
\eeq
In practice the spectral shape of the resulting $\bar p$ fluxes from all these channels are not very different, due to the qualitative fact that $\bar p$ are the final product of a complex hadronization process which `washes out' possible distinguishing features. This fact, we anticipate, will imply that the channel cannot be easily determined on the basis of data fitting (see sec.~\ref{AMSfit}). \\
Moreover, some of the channels yield $\bar p$ spectra which are practically indistinguishable, both in terms of shape and normalization. This is the case for $b \bar b$ and $t \bar t$ (of course for the range of DM masses in which both are kinematically allowed), for $W^+W^-$ and $ZZ$ as well as for $\mu^+ \mu^-$ and $\tau^+ \tau^-$). Therefore, for all practical purposes, only four of the channels lead to actually different results ($b \bar b, W^+W^-, \mu^+ \mu^-$ and $\gamma \gamma$) and thus we will mainly focus on these in the following. Along these same lines of argument, the results for other channels that we do not consider can be safely deduced from channels with similar spectra: for instance, the $hh$ channel (where $h$ denotes the $\sim$ 125 GeV SM higgs boson) is equivalent to the $b \bar b$ one, the $e^+e^-$ is equivalent to the $\mu^+ \mu^-$ one etcetera.

Notice that, alongside with the `traditional' quark and weak gauge boson channels (collectively: `hadronic channels'), we consider leptonic channels ($\mu^+ \mu^-, \tau^+ \tau^-$) and a direct $\gamma\gamma$ channel. Antiprotons are produced in these channels because we take into account electromagnetic and electroweak corrections (i.e. splitting of the initial particles and radiation of electro(weak) bosons) to the tree level production process. Not surprisingly, the yield of $\bar p$ from these processes is suppressed and therefore (we anticipate) the resulting constraints will be much weaker than for hadronic channels. Nevertheless, we include these channels for completeness and for the interest that they have carried in recent years for the interpretation in terms of DM of the \PAMELA\ and \Fermi\ $e^\pm$ anomalies and of the 135 GeV line in \Fermi\ data.

\medskip

The antiprotons produced by annihilations or decays in any given point of the galactic DM halo are then subject to propagation in the galactic environment. The prediction for the fluxes collected at Earth depends therefore on the assumed distribution of DM in the halo and on the assumed propagation parameters.  
Concerning the DM halo profiles we consider three of the usual possible choices motivated by numerical N-body simulations or by observations, namely:
\begin{equation}
\begin{array}{rrcl}
{\rm NFW:} & \rho_{\rm NFW}(r)  & = & \displaystyle \rho_{s}\frac{r_{s}}{r}\left(1+\frac{r}{r_{s}}\right)^{-2}, \\[4mm]
{\rm Einasto:} & \rho_{\rm Ein}(r)  & = & \displaystyle \rho_{s}\exp\left\{-\frac{2}{\alpha}\left[\left(\frac{r}{r_{s}}\right)^{\alpha}-1\right]\right\}, \\[4mm]
{\rm Burkert:} & \rho_{\rm Bur}(r) & = & \displaystyle  \frac{\rho_{s}}{(1+r/r_{s})(1+ (r/r_{s})^{2})}. \\[4mm]
\end{array}
\label{eq:profiles}
\end{equation} 
The Einasto~\cite{Graham:2005xx, Navarro:2008kc} and NFW profiles~\cite{Navarro:1995iw} (favored by simulations) are peaked towards the Galactic Center while the Burkert profile~\cite{Burkert} (possibly more in agreement with observations of galactic rotation curves) sports a constant-density core at the center. The specific parameters to be plugged in eq.~(\ref{eq:profiles}) are given in Table~\ref{tab:proparam}.

Concerning the propagation parameters, we adopt the standard three sets reported in Table~\ref{tab:proparam} (right), derived from consistency with ordinary Cosmic Ray (CR) measurements. These parameters are to be plugged into the equation that governs the diffusion-loss propagation of antiprotons in the galactic environment: the set denoted by `MIN' (`MAX') corresponds to a minimal (maximal) final yield of $\bar p$ and therefore they bracket the uncertainty. We will briefly comment later on how favored each set should be considered. 

\begin{table}[t]
\parbox{.45 \linewidth}{\small
\begin{tabular}{l|crc}
 &  \multicolumn{3}{c}{DM halo parameters}  \\
DM halo & $\alpha$ &  $r_{s}$ [kpc] & $\rho_{s}$ [GeV/cm$^{3}$]\\
  \hline 
  NFW & $-$ & 24.42 & 0.184 \\
  Einasto & 0.17 & 28.44 & 0.033 \\
  Burkert & $-$ & 12.67 & 0.712 \\
 \end{tabular} 
 } \
 \parbox{.45\linewidth}{\small
\begin{tabular}{c|cccc}
 &  \multicolumn{4}{c}{Antiproton propagation parameters}  \\
Model  & $\delta$ & $\mathcal{K}_0$ [kpc$^2$/Myr] & $V_{\rm conv}$ [km/s] & $L$ [kpc]  \\
\hline 
MIN  &  0.85 &  0.0016 & 13.5 & 1 \\
MED &  0.70 &  0.0112 & 12 & 4  \\
MAX  &  0.46 &  0.0765 & 5 & 15 
\end{tabular}
}
\caption{\em \small {\bfseries Astrophysical parameters.} Left panel: the parameters for the considered DM profiles, to be plugged in the functional forms of eq.~(\ref{eq:profiles}). These specific values are derived as discussed in~\cite{PPPC4DMID}. Right panel: propagation parameters for anti-protons in the galactic halo (from~\cite{FornengoDec2007,DonatoPRD69}). Here $\delta$ and $\mathcal{K}_0$ are the index and the normalization of the diffusion coefficient, $V_{\rm conv}$ is the velocity of the convective wind and $L$ is the thickness of the diffusive cylinder. 
\label{tab:proparam}}
\end{table}

\medskip

Concretely, we obtain the $\bar p$ fluxes at Earth (post-propagation) from the numerical products provided in~\cite{PPPC4DMID}. These fully take into account the EW corrections mentioned above~\cite{Ciafaloni:2010ti}. 
As an example, the $\bar p$ flux from a model we will consider below is plotted in fig.~\ref{fig:flux_data}, right.
During the whole analysis, only kinetic energies above 10 GeV are considered, to minimize the effect of solar modulation. 

\subsection{Secondaries}
\label{secondaries}

The term `secondary antiprotons' refers to the smooth $\bar p$ flux produced by standard CR processes (essentially the collisions of primary $p$ on the interstellar medium), which constitutes the background to the DM signal. 

\begin{figure}[t]
\begin{center}
\includegraphics[width=0.47\textwidth]{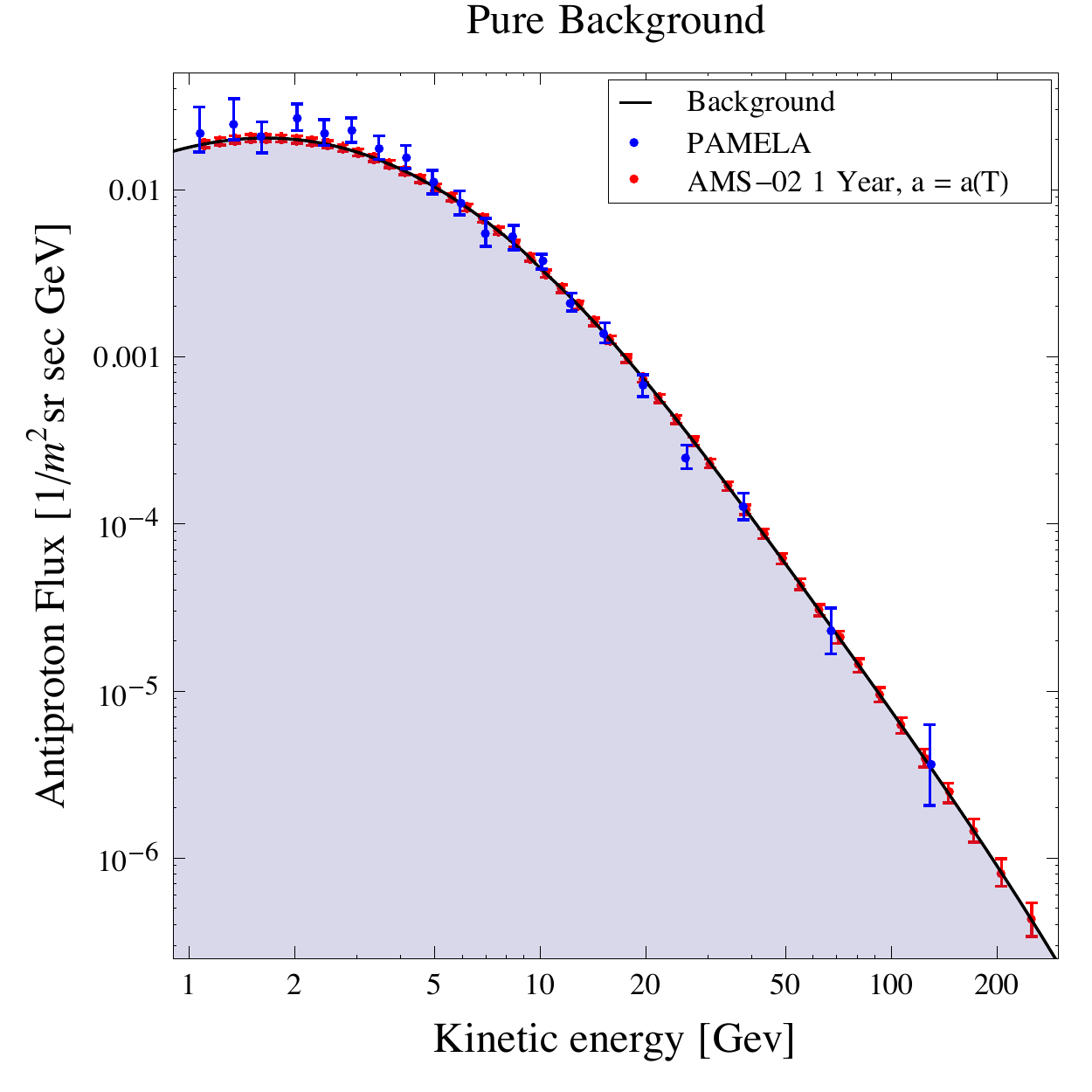}\quad
\includegraphics[width=0.47\textwidth]{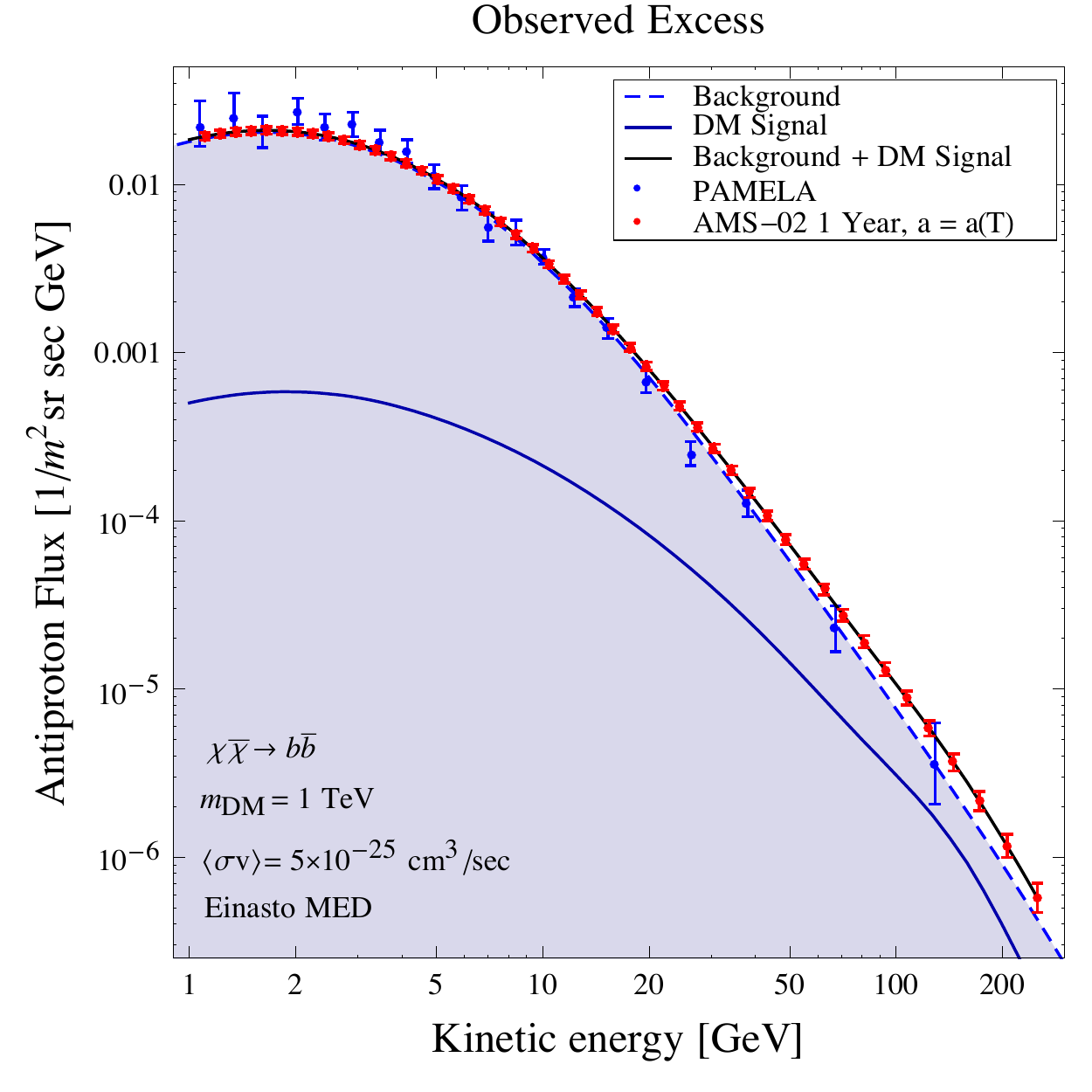}
\caption{\small \em\label{fig:flux_data} {\bf Examples of antiproton fluxes and data.} {\em Left Panel:} astrophysical $\bar p$ background, superimposed to the current data from \PAMELA\ and to one of our realizations of mock data from the \AMS\  experiment. {\em Right Panel:} the same but including a contribution from DM annihilation giving rise to an excess on top of the background which could be detected by \AMS. The example chosen is DM annihilating into $b\bar{b}$ with $m_{\DM}=1$ {\em TeV} and $\langle \sigma v\rangle = 5\times 10^{-25}$ {\rm cm}$^3${\rm s}$^{-1}$ for an Einasto halo profile and `MED' propagation parameters.}
\end{center}
\end{figure}

We do not calculate such background ourselves, but use instead the recent determination obtained by~\cite{Evoli:2011id}. There, the two extreme cases are the Kolmogorov (KOL) and Kraichanian (KRA) models: we choose a spectrum which goes roughly through the middle of these (for illustration, it is the one plotted in fig.~\ref{fig:flux_data}). We will refer to this in the following as the `fixed background'.

\medskip

We then have to model the uncertainty of such background function: we let the amplitude vary by 10\% and the spectral index  by $\pm$0.05. This envelops fairly generously the different models and the astrophysical uncertainties. Technically, we just multiply the fixed background by $A \, T^p$ (where $T$ is the antiproton kinetic energy), and allow $A \in [0.9, 1.1]$ and $p \in [-0.05, 0.05]$. In the following, we will marginalize with respect to the $A$ and $p$ parameters, i.e. we will identify the values that give the best fit for a given assumed signal and a given set of data. We refer to this procedure as `varying background' (see the detailed discussions below). In absence of a signal, the best fit background with respect to the \PAMELA\ data is obtained with an amplitude of $A_0=1.08$ and a slope $p_0=-0.05$: it gives a $\chi_0^2(A_0, p_0) = 4.62$ for 8 data points (we always restrict to kinetic energies $T>10$ GeV).

We note that, in previous analyses (e.g.~\cite{Belanger:2012ta}), the background used was the one in~\cite{Bringmann:2006im} with wider intervals for the variating amplitude (40\%) and slope ($\pm 0.10$). The obtained uncertainty band was thus broader and these choices would allow to fit the \PAMELA\ data slightly better. However, we decide to make use of the function deduced from~\cite{Evoli:2011id} since this analysis takes recent cosmic ray and proton spectra into account and the uncertainties are clearly smaller. It is of course possible to get the same larger bandwidth from the new background by increasing the amplitude and the slope.

\medskip

Before moving on, we comment on the robustness of the assumption that the background follows a smooth shape, essentially consisting of a power law (above 10 GeV). While this is expected to follow from most astrophysical processes, it may not always be strictly the case. In fact, some astrophysical sources, such as supernova remnants~\cite{Blasi:2009bd} or plasma phenomena~\cite{Blasi:2012yr}, could induce a different spectrum. However, such deviations from an effective power law are typically expected at energies higher than those of our interest~\cite{Donato:2010vm}. Nevertheless, in case of a positive detection of a deviation, it will be crucial to keep in mind these possibilities when working on its interpretation.

\section{Results and discussion}
\label{results}

\subsection{Current antiproton constraints from \PAMELA}
\label{PAMELAbound}

The currently most precise measurement of the CR antiproton flux is provided by the \PAMELA\ satellite~\cite{Adriani:2010rc}, as anticipated in the Introduction. The data, reproduced in fig.~\ref{fig:flux_data}, extend from kinetic energies of less than 1 GeV to about 180 GeV (although we use only the portion above 10 GeV to avoid the effects of solar modulation). 

\begin{figure}[t]
\begin{center}
\includegraphics[width=0.47\textwidth]{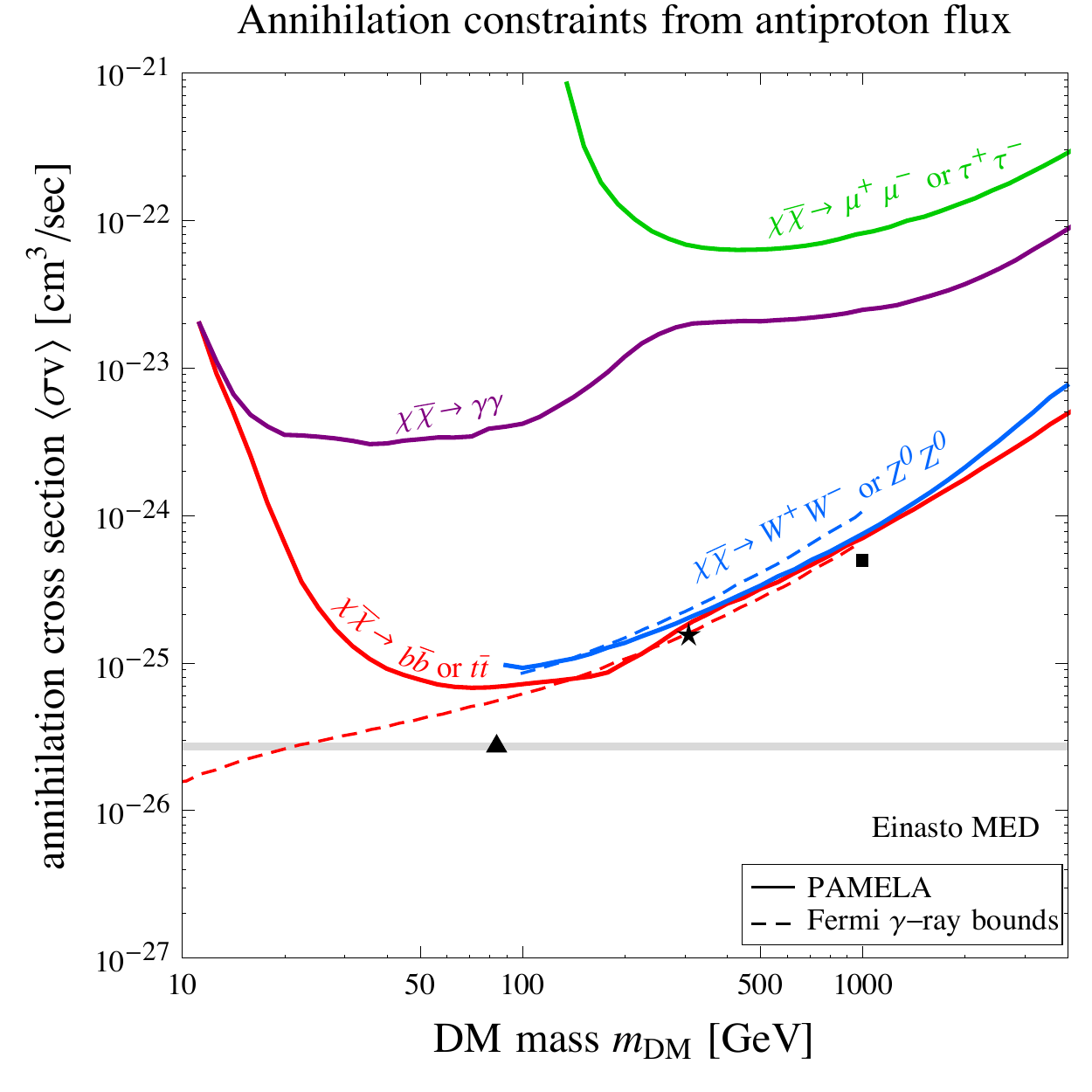}\quad
\includegraphics[width=0.47\textwidth]{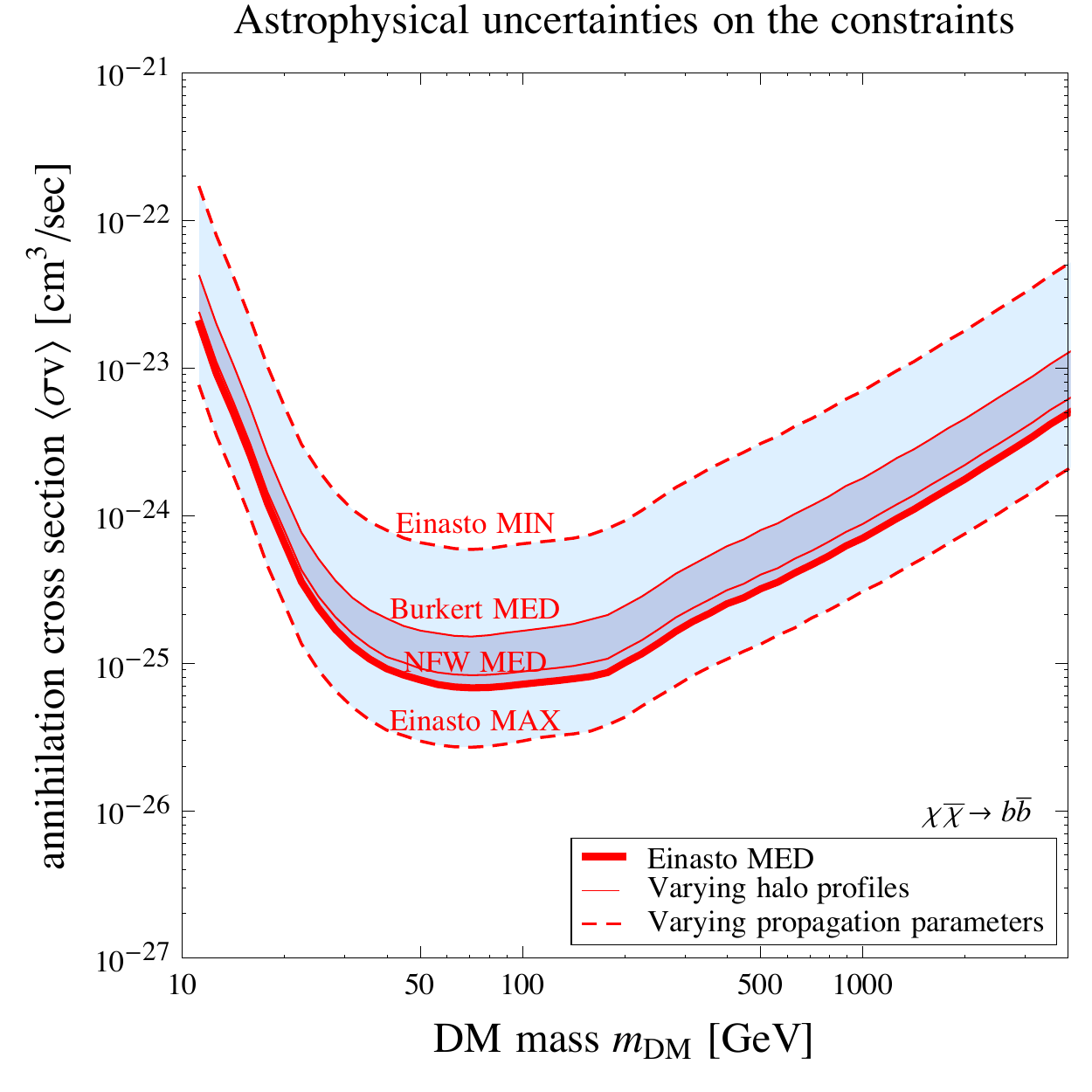}
\caption{\small \em\label{fig:constraints_ann} {\bf Annihilating DM: current constraints.} {\em Left Panel:} current constraints from the antiproton measurements by \PAMELA, for different annihilation channels. The areas above the curves are excluded. The dashed lines reproduce the $\gamma$-ray constraints from~\cite{Ackermann:2011wa}, for the same channels. The symbols individuates the parameters used for the analyses in Sec.~\ref{AMSfit} while the horizontal band signals the thermal relic cross section. {\em Right Panel:} illustration of the impact of astrophysical uncertainties: the constraint for the $b \bar b$ channel spans the shaded band when varying the propagation parameters (dashed lines) or the halo profiles (solid lines).}
\end{center}
\end{figure}

\begin{figure}[t]
\begin{center}
\includegraphics[width=0.47\textwidth]{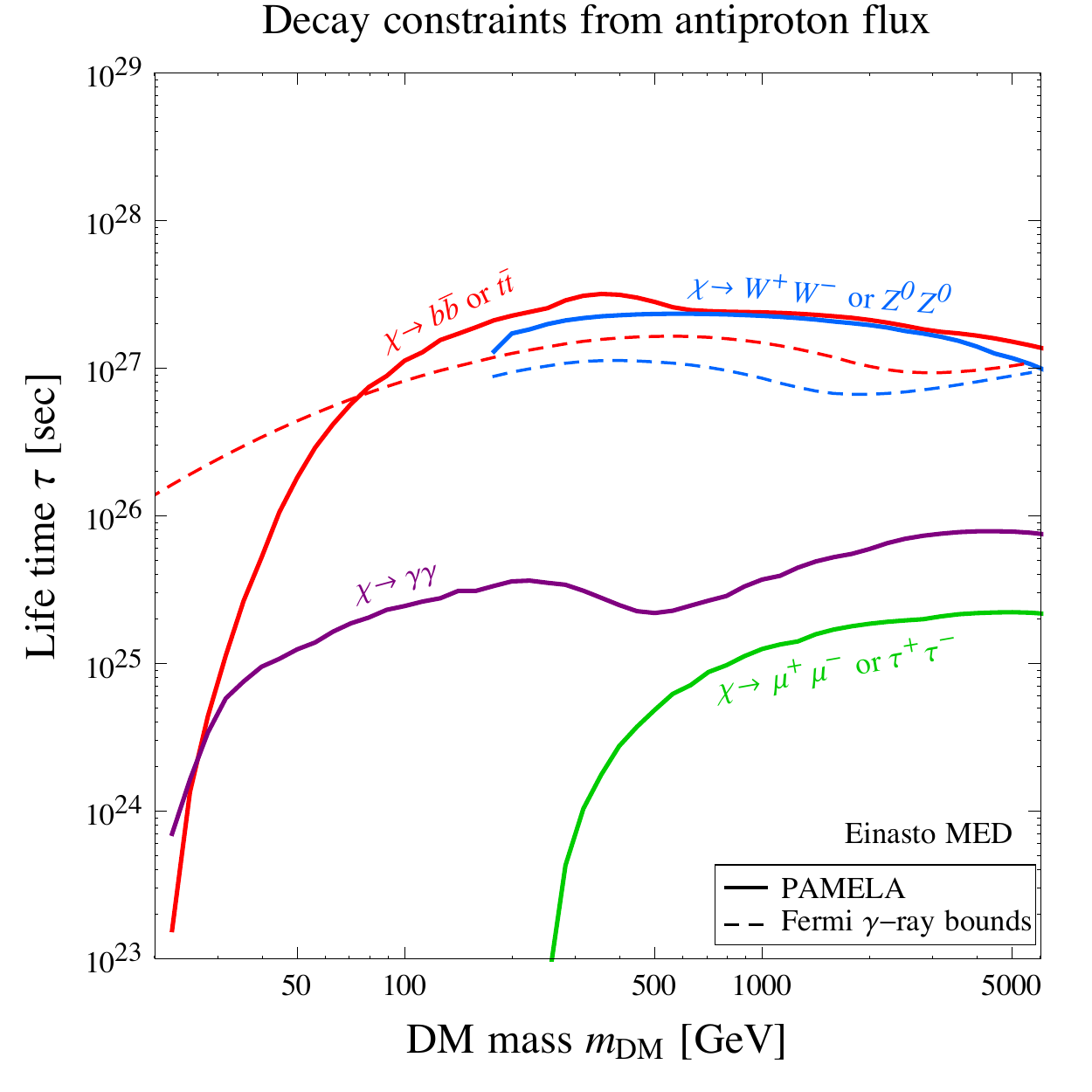}\quad
\includegraphics[width=0.47\textwidth]{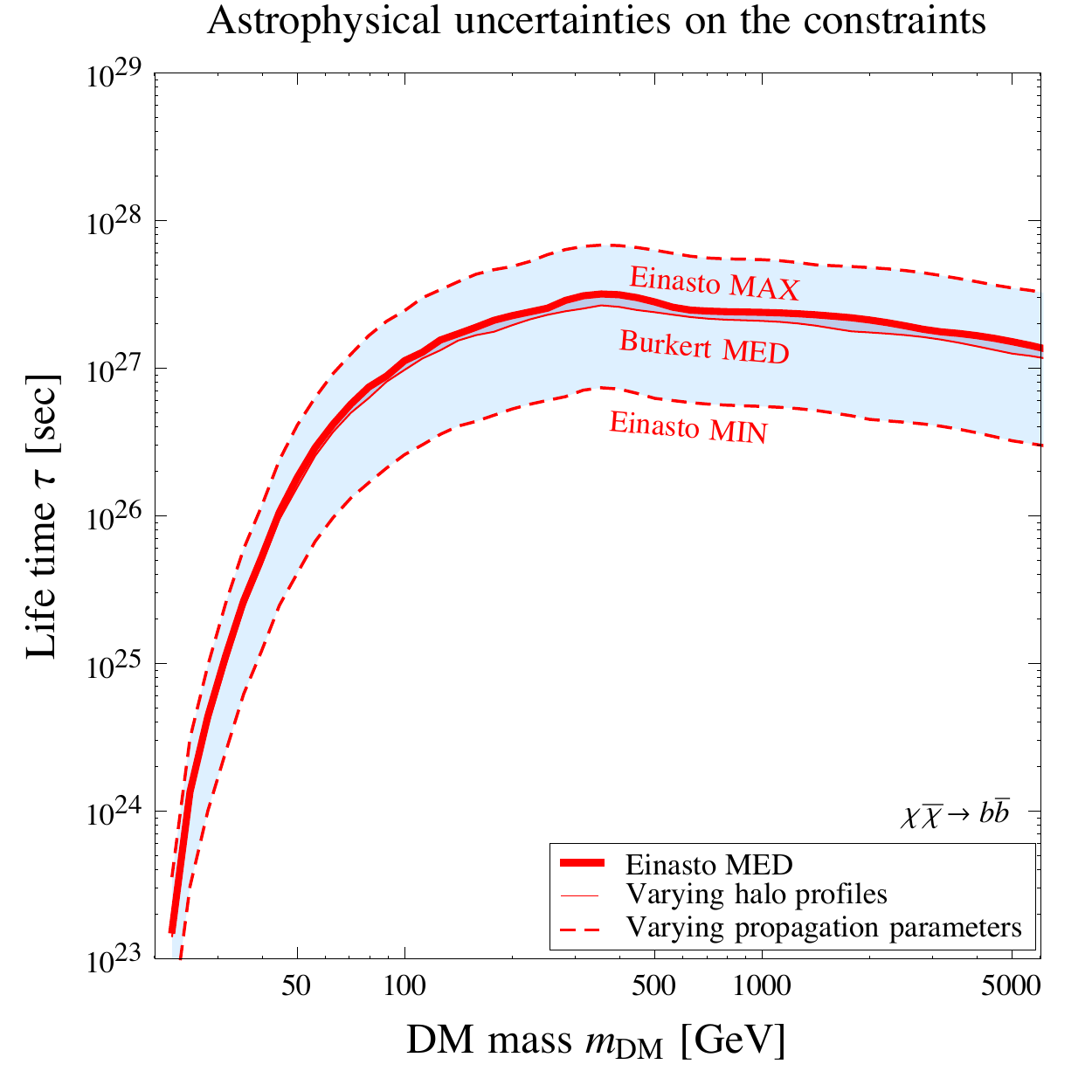}
\caption{\small \em\label{fig:constraints_decay} {\bf Decaying DM: current constraints.} {\em Left Panel:} current constraints from the antiproton measurements by \PAMELA, for different decay channels. The areas below the curves are excluded. The dashed lines reproduce the $\gamma$-ray constraints from~\cite{Cirelli:2012ut}. {\em Right Panel:} illustration of the impact of astrophysical uncertainties: the constraint for the $b \bar b$ channel spans the shaded band when varying the propagation parameters (dashed lines) or the halo profiles (solid lines).}
\end{center}
\end{figure}

The total antiproton flux is given by the sum of the DM and the astrophysical contributions,
\beq
\phi_{\rm tot}(m_{\rm DM}, \langle \sigma v \rangle; A,p)= \phi_{\rm DM}(m_{\rm DM}, \langle \sigma v \rangle)+ \phi_{\rm bkg}(A,p).
\eeq
For fixed values of the DM particle mass $m_{\rm DM}$ and the thermally averaged annihilation cross-section $\langle \sigma v \rangle$, the astrophysical background is optimized within the uncertainty bandwidth in order to minimize the $\chi^2$ of the total flux with respect to the data. During this procedure, the optimal values of the amplitude $A_{\rm opt} \in [0.9,1.1]$ and the slope $p_{\rm opt} \in [-0.05,0.05]$ are determined. Then, to find the 2$\sigma$ exclusion contour in the ($m_{\rm DM}, \langle \sigma v\rangle$)-plane, the required condition is 
\beq
\Delta \chi^2(m_{\rm DM}, \langle \sigma v\rangle)= \chi^2(m_{\rm DM}, \langle \sigma v\rangle ; A_{\rm opt}, p_{\rm opt})-  \chi_0^2(A_0,p_0)< 4.
\eeq

We show the resulting contours in figures~\ref{fig:constraints_ann} (for annihilating DM) and~\ref{fig:constraints_decay} (for decaying DM). In the left panels we have fixed the astrophysical parameters to a fiducial choice: the halo model is chosen to be Einasto and the propagation parameters to be `MED'. The constraints on $\langle \sigma v \rangle$ reach the level of $\sim 10^{-25}$ cm$^3$/sec for the quark or gauge boson channels, at a DM mass of about 100 GeV. For decaying DM, the constraints on the lifetime $\tau$ reach $\sim$ few $\times \ 10^{27}$ sec. The bounds for the exotic channels ($\gamma\gamma$ and leptons) are much less stringent, as expected. 
The constraint from the leptonic channels is lifted below $m_{\rm DM} \simeq 300$ GeV, consistently with the fact that EW corrections are effective only for large $m_{\rm DM}$~\cite{Ciafaloni:2010ti}. The $\gamma\gamma$ constraint exhibits two regimes: at $m_{\rm DM} \lesssim 300$ GeV some $\bar p$ are produced by $\gamma$ showering; at $m_{\rm DM} \gtrsim 300$ GeV EW corrections set in and increase the $\bar p$ yield, resulting in a constraint that follows the shape of the leptonic one.

\medskip

These results show that, for the hadronic channels, the constraints inferred from the data of \PAMELA\ are already competitive with (or even slightly better than) the bounds from \Fermi\ $\gamma$-rays (reported as dashed lines in figures~\ref{fig:constraints_ann} and~\ref{fig:constraints_decay}, respectively from~\cite{Ackermann:2011wa}\footnote{See also~\cite{othergammadwarfs}.} and~\cite{Cirelli:2012ut}), for both annihilating and decaying DM. This is a non-trivial conclusion, especially considering that the prospects of improvement in $\gamma$-rays are arguably less promising than in antiprotons (the \Fermi\ satellite will at most double the statistics in the final years of its lifetime, while for antiprotons \AMS\ can in principle accumulate high precision data for several years to come). 

\medskip

On the other hand, the astrophysical uncertainties for antiprotons are non-negligible: the choice of propagation parameters and halo profile can change the constraints by one or two orders of magnitude, as illustrated in the right panels of fig.s~\ref{fig:constraints_ann} and~\ref{fig:constraints_decay}.
The impact of changing the propagation parameters from `MIN' to `MAX' is particularly important as it accounts by itself for more than a one order of magnitude shift in the bounds.\footnote{It is worth noticing, however,  that the propagation model `MIN' has already been disfavored due to its small thickness of the diffusion cylinder. Studies on synchrotron emission, radio maps and low energy positron spectrum exclude a diffusion halo scale smaller than 2 kpc at 3$\sigma$ level~\cite{DiBernardo:2012zu}. Moreover, a bayesian analysis on B/C, $^{10}$Be/$^9$Be ratios and carbon and oxygen spectra suggests a thickness in the interval $[3.2, 8.6]$ kpc at 95\% confidence level \cite{Trotta:2010mx}. See also the analysis in~\cite{Ackermann:2012rg}, whose results also disfavors a 1 kpc thickness, although the observation region is not optimized and the results are weakened by the inclusion of Dark Matter.} We also notice that, as expected, the impact of changing the halo profile is more important for the annihilation case than for the decay one.  In fact, the differential $\bar p$ flux is sensitive to the halo density {\em squared} for annihilations while it is to the first power in the case of decay.

\subsection{Projected sensitivity of \AMS}
\label{AMS}

In this section we move to study the reach of the upcoming early \AMS\ antiproton results on DM.

\medskip

We will therefore need to produce mock \AMS\ data and their associated error bars, which we do by implementing the following procedure. We adopt a linear approximation of the rigidity resolution of \AMS\ as given in \cite{TingTalk}:
\begin{equation}
r(T)=\frac{\Delta T}{T}=0.0042\times T+0.1,
\end{equation}
where $T$ is the kinetic energy of the incoming antiproton. This allows us to determine a realistic $\bar p$ energy binning.
We assume that the $\bar p$ flux will be measured up to 300 GeV (we will use only the portion above 10 GeV, as already discussed, for a total of 27 simulated data points), thus extending the range of \PAMELA\ to higher energies. 
Next, we need to estimate the errors on the measurement of the antiproton flux. The number of collected and reconstructed antiprotons in \AMS\ in a given bin $i$ centered around a kinetic energy $T_i$ will be given by
\begin{equation} 
N_i= \epsilon \, a(T_i) \, \phi(T_i) \, \Delta T_i \, \Delta t,
\end{equation}
where $\epsilon$ is the efficiency, $a$ the geometrical acceptance of the apparatus (which is in general a function of the energy), $\phi$ the antiproton flux, $\Delta T$ the width of the kinetic energy bin and $\Delta t$ the exposure time. Assuming a Poisson distribution ($\Delta N= \sqrt{N}$), the statistical error on the antiproton flux in bin $i$ is therefore
\begin{equation}
\Delta \phi_i \left|_{stat}\right.= \sqrt{\frac{\phi(T_i)}{\epsilon \, a(T_i) \, \Delta T_i \, \Delta t}}.
\end{equation}
We assume systematic errors of 5\% based on the analysis in~\cite{Pato:2010ih} and we sum them in quadrature with the statistical ones. The $\bar p$ efficiencies of the different subdetectors of \AMS\ are expected to be close to 1 (or in any case well above 0.9) for the whole range of energies of our interest~\cite{Choutko}. Therefore we set the overall efficiency $\epsilon=1$. The modelization of the acceptance of the experiment is the most critical point and the one which is least accessible from outside the collaboration. We will therefore consider two options: 
\begin{enumerate} 
\item \label{case1}  $\Delta t=1$ year and an acceptance $a=0.147$ m$^2$ sr for $1$ GeV $<T<11$ GeV, but an energy dependent one (contained roughly in the interval $a \sim 0.02 \div 0.04$ m$^2$ sr) for energies $T>11$ GeV, by interpolating the points in fig. 8 of~\cite{Malinin:2004pw};
\item \label{case2} $\Delta t=3$ years with a constant acceptance for all energies $a = 0.1$ m$^2$ sr.
\end{enumerate}
The second case is obviously much more optimistic, as it gives rise to a much larger accumulated statistics. The first case might be more realistic though: in \AMS, the identification of antiprotons at high energy will likely have to rely on the use of the calorimeter, thus reducing the pure geometrical acceptance of the experiment.
We can expect the real data to lie in between these two scenarios. 
The resulting set of mock data (for option 1, for definiteness) is reproduced in fig.~\ref{fig:flux_data}, both for a scenario with no excess (left) and for a scenario where an excess over background is detected (right). We indeed now move to discuss these two cases.

\subsubsection{Foreseen constraints}
\label{AMSbound}

First, we consider a scenario in which the measurements of \AMS\ do not shown an antiproton excess and the data points follow the background, as represented on Figure \ref{fig:flux_data} on the left. 

\begin{figure}[t]
\begin{center}
\includegraphics[width=0.47\textwidth]{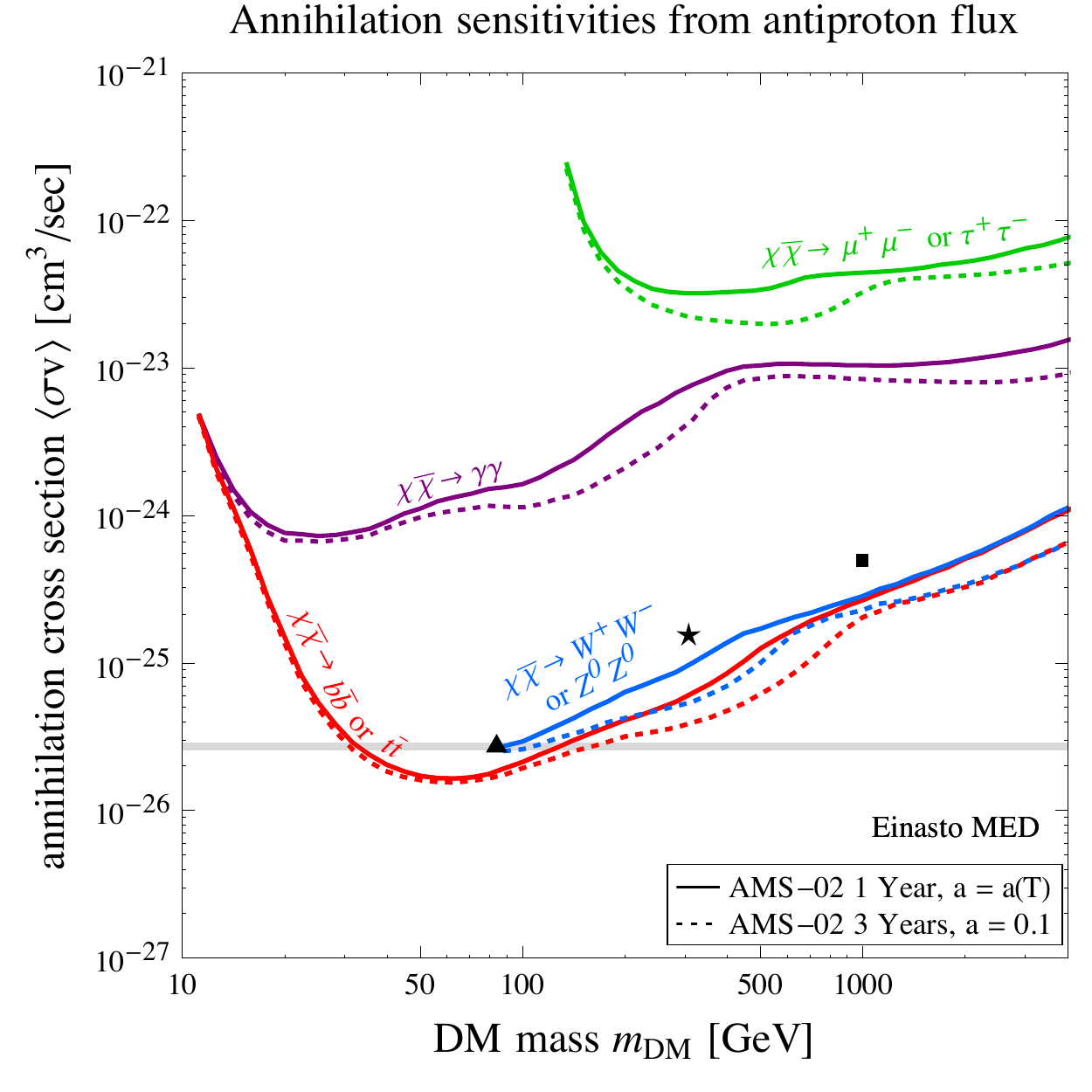}\quad
\includegraphics[width=0.47\textwidth]{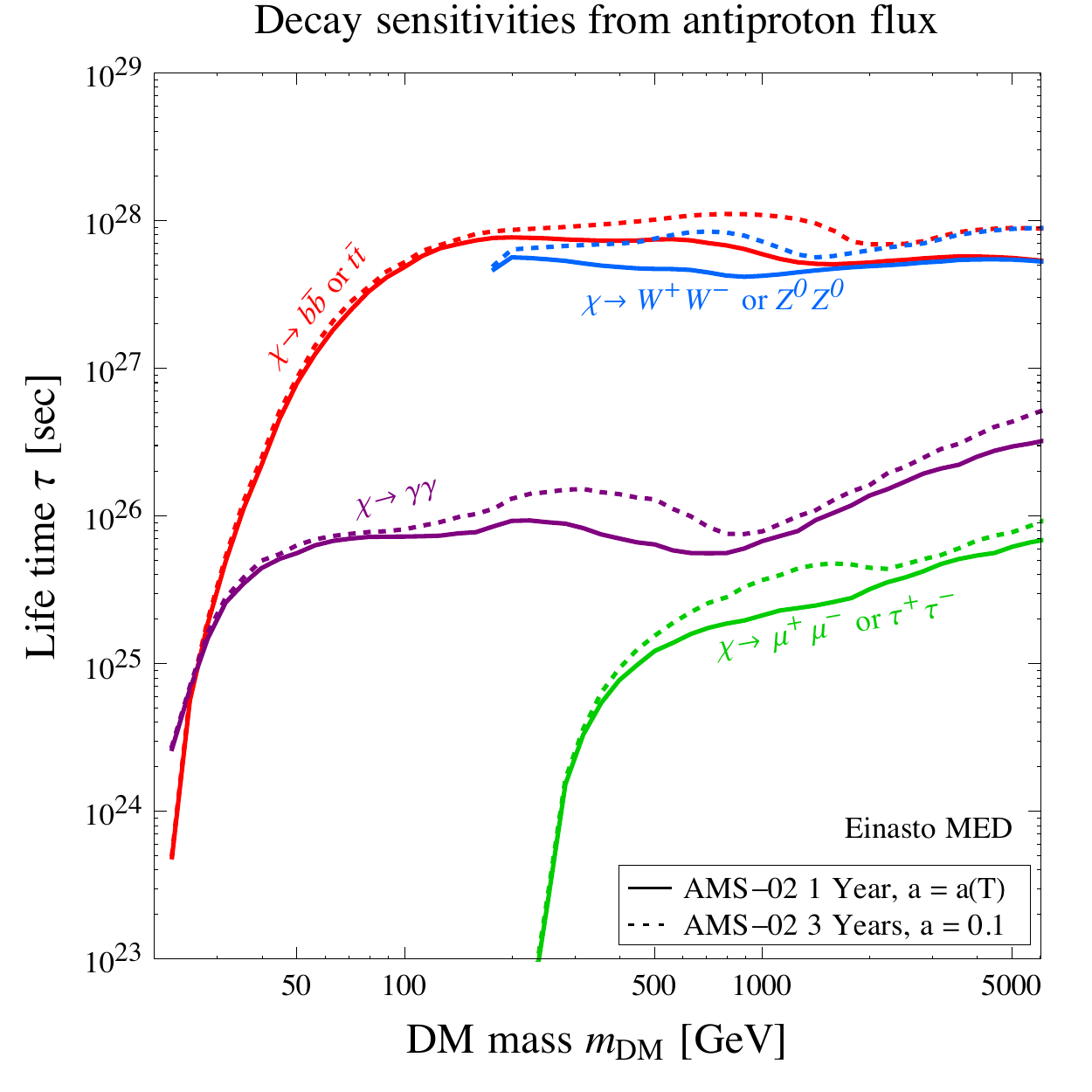}
\caption{\small \em\label{fig:AMS_ann_decay} {\bf Annihilating and decaying Dark Matter: future sensitivities.} {\em Left panel:} the sensitivity of early \AMS\ antiproton measurements for different channels. The solid lines assume 1 year of data taking and a realistic acceptance, the dotted lines assume 3 years of data taking and an optimized acceptance. The rest of the notations are like in fig.~\ref{fig:constraints_ann}. {\em Right panel:} the same for the case of decaying DM.}
\end{center}
\end{figure}

We generate the mock data using the `fixed background' spectrum function. We then compute the constraints for annihilation and decay of DM with the same method discussed in Sec.~\ref{PAMELAbound} (therefore, in particular, implementing the `varying background' procedure). We believe this `workflow' to be realistically similar to an actual data analysis. The results are presented in Figure \ref{fig:AMS_ann_decay}.

\medskip

It is apparent that, with \AMS, the constraints can be improved by about an order of magnitude or slightly less with respect to \PAMELA. For the $b \bar b$ channel (and for the $t \bar t$ one, when kinematically open) the bounds will reach the thermal value of the annihilation cross-section ($2.7 \times 10^{-26}$ cm$^3$/sec) in the range of masses 30 $-$ 200 GeV. 
The more optimistic choice for the acceptance and the operating time (option~\ref{case2} in~\ref{AMS}) improves the projected reach by at most another half an order of magnitude or so, except at low masses (i.e. low energies) where the systematic errors dominate.

\medskip

We only show the curves corresponding to the `Einasto, MED' choice of astrophysical parameters: other choices would modify the bounds in the same way we discussed in Sec.~\ref{PAMELAbound}. However, we also mention that a significant collateral improvement expected from \AMS\ will be to reduce the astrophysical uncertainties: using low energy spectra and the B/C ratio, as described above~\cite{DiBernardo:2012zu, Trotta:2010mx}, the propagation of cosmic rays should be better pinned down and the error band reduced.

\medskip

In~\cite{Evoli:2011id} a similar analysis has been performed, for annihilating DM. Our results in fig.~\ref{fig:AMS_ann_decay} (left) compare almost directly with their fig. 17, albeit on a reduced range of masses and for the specific models that they consider. The constraints that we find are less stringent than theirs by about one order of magnitude. We attribute this to the fact that they assume a larger acceptance and, possibly, to their use of a larger range of energies.

\subsubsection{Reconstruction capabilities}
\label{AMSfit}

We now consider a scenario in which an identifiable excess is detected by \AMS\ in the antiproton flux and it is attributed to DM, and we study how well it will be possible to reconstruct the underlying DM properties. We focus only on annihiliating DM, for definiteness, in this Subsection. 

\begin{figure}[p]
\begin{center}
\includegraphics[width=0.32\textwidth]{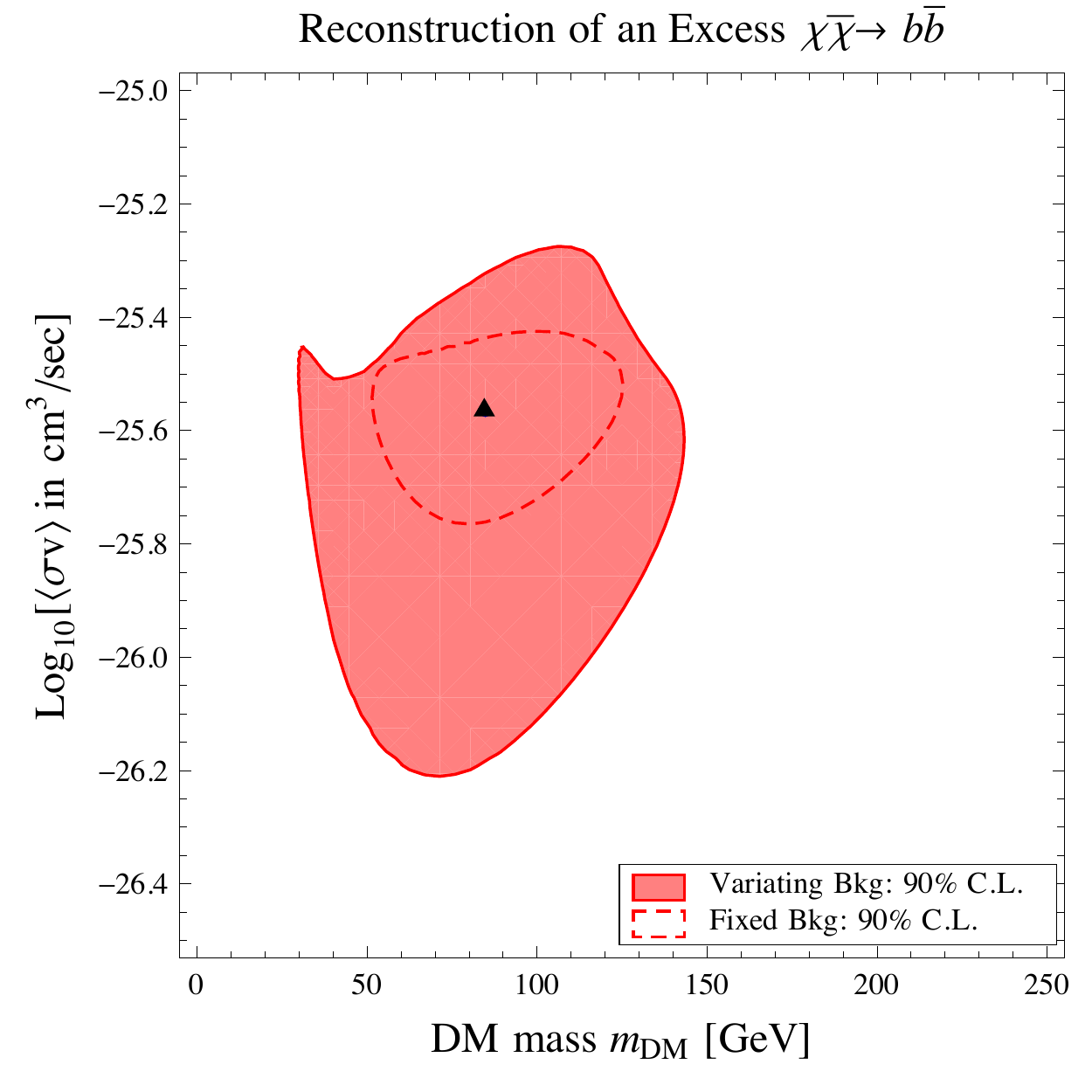}\ \
\includegraphics[width=0.32\textwidth]{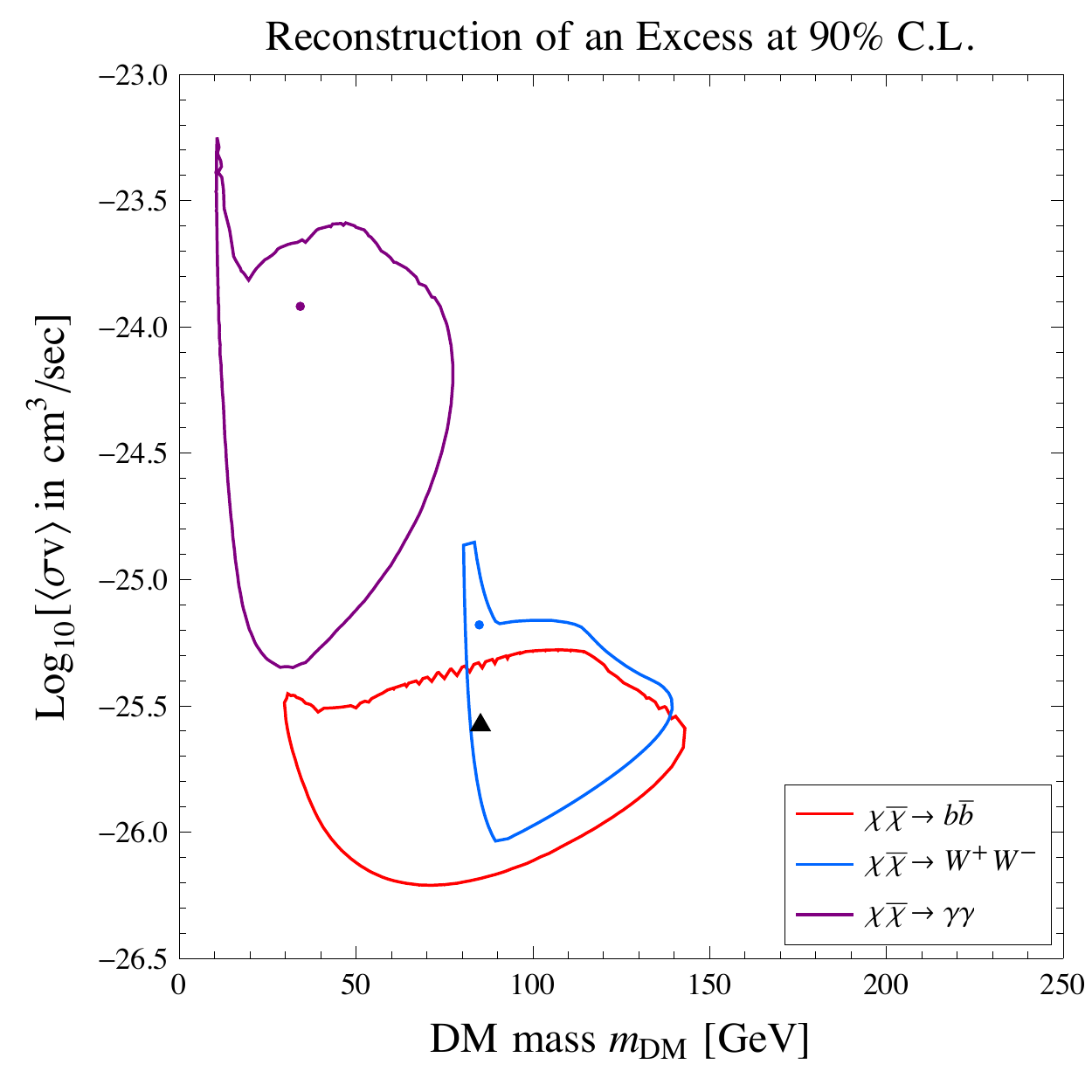}\ \
\includegraphics[width=0.32\textwidth]{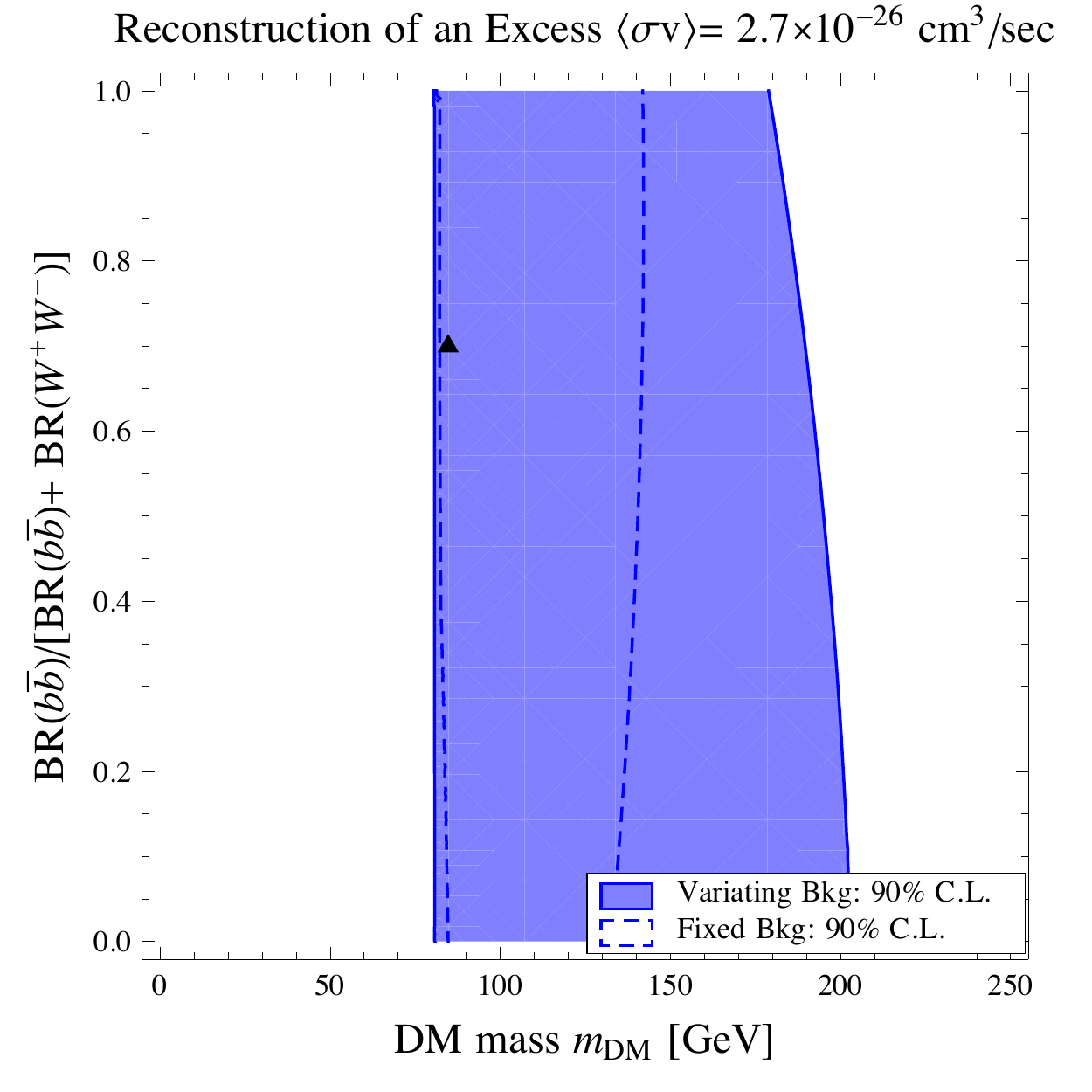}\\
\includegraphics[width=0.32\textwidth]{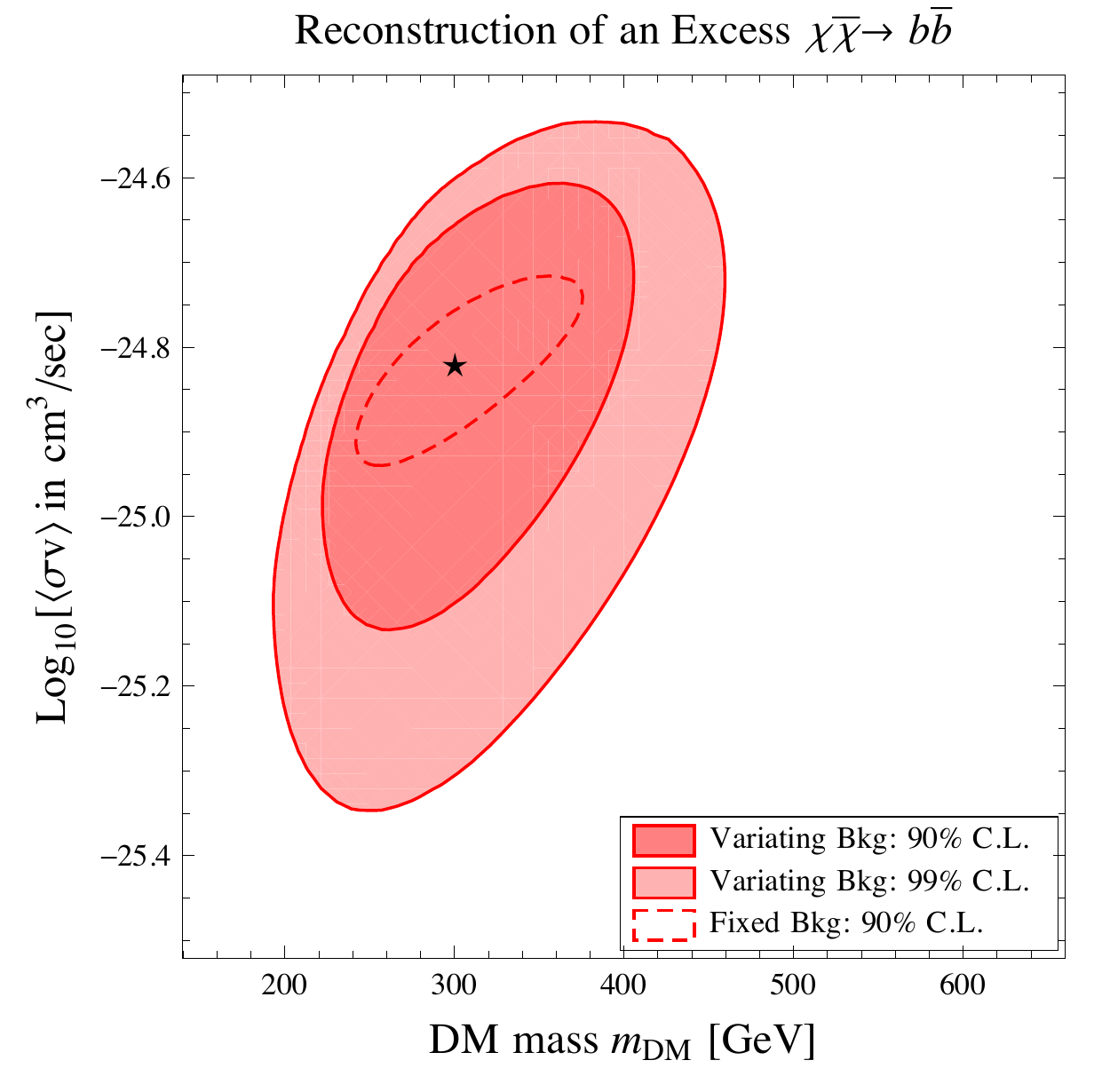}\ \
\includegraphics[width=0.32\textwidth]{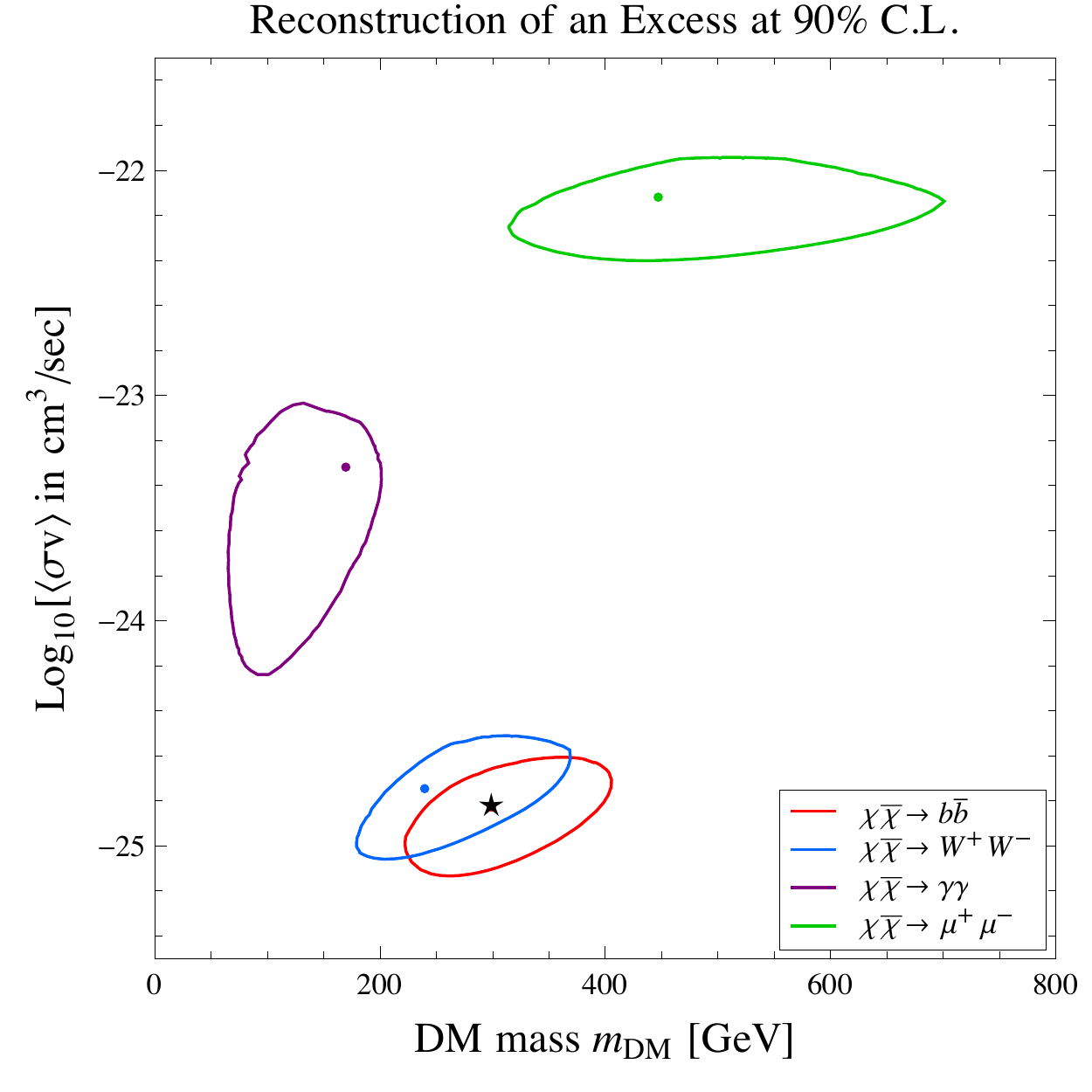}\ \
\includegraphics[width=0.32\textwidth]{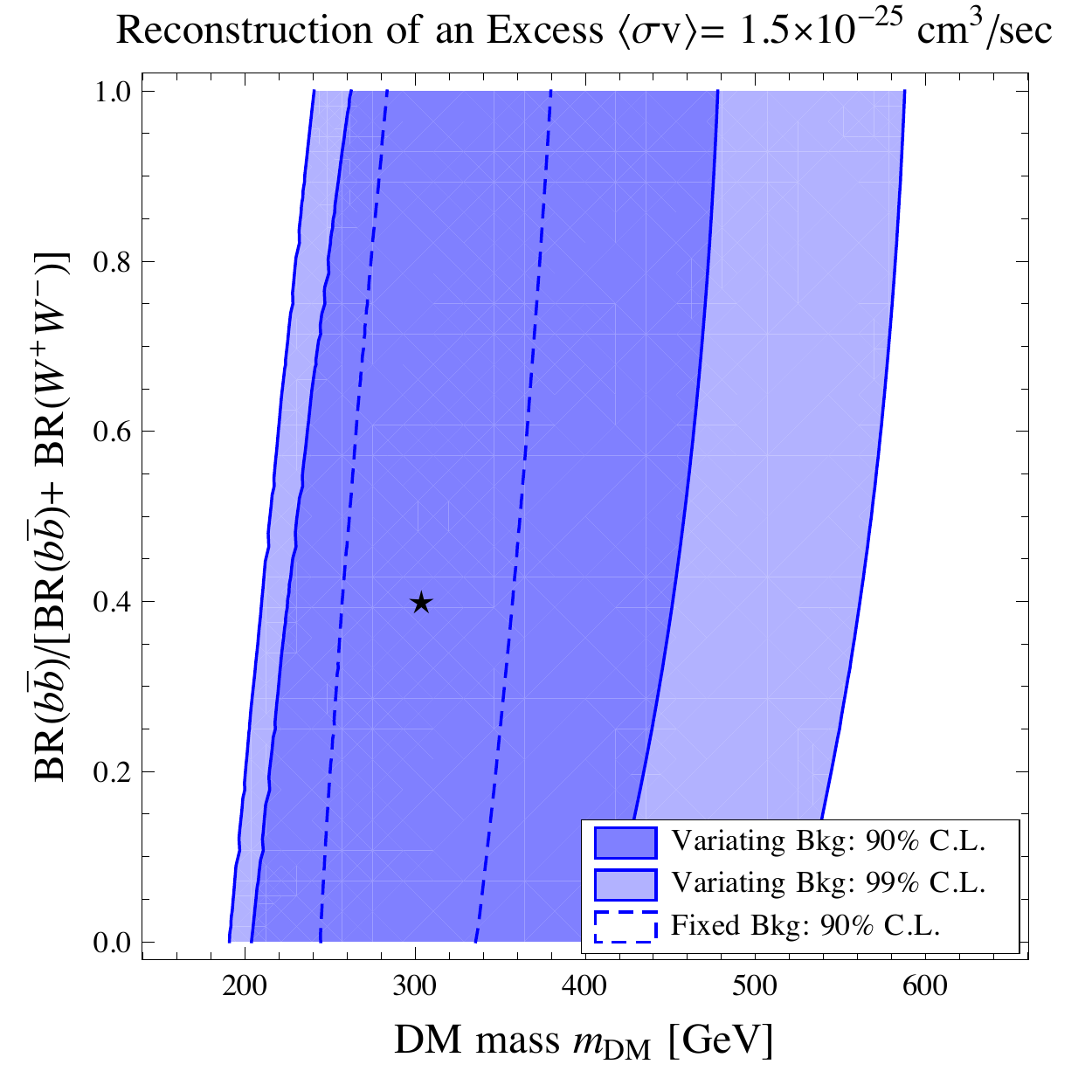}\\
\includegraphics[width=0.32\textwidth]{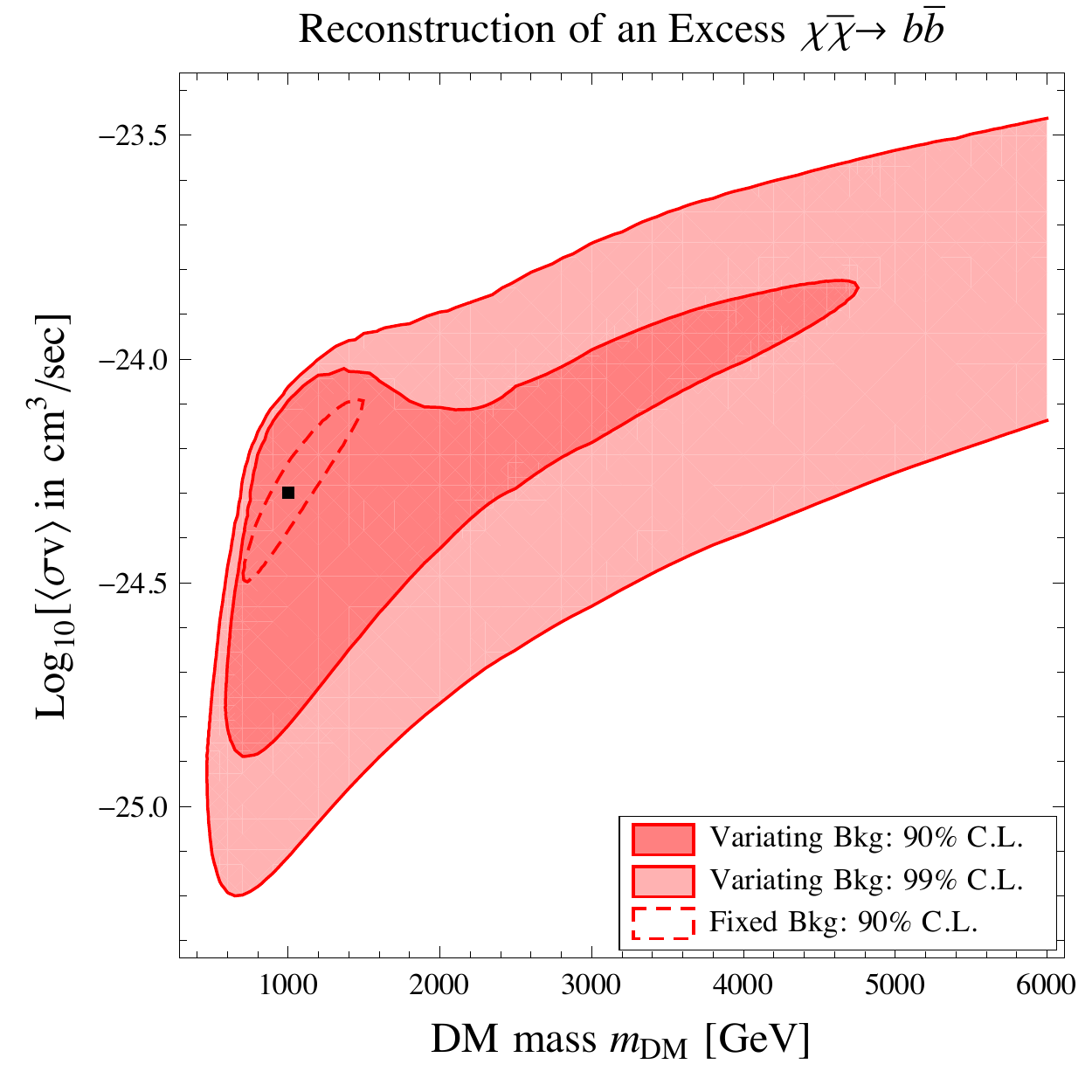}\ \
\includegraphics[width=0.32\textwidth]{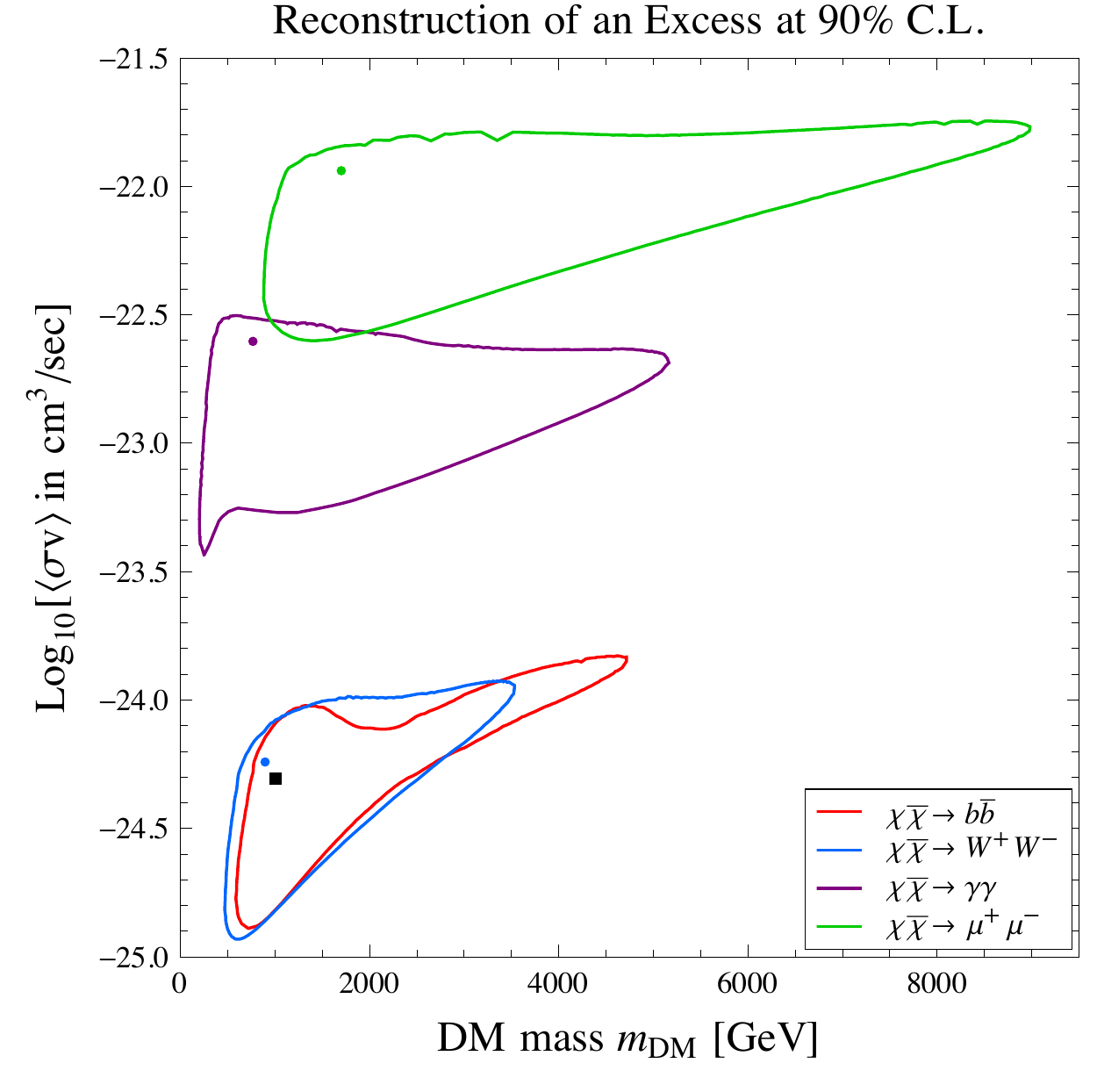}\ \
\includegraphics[width=0.32\textwidth]{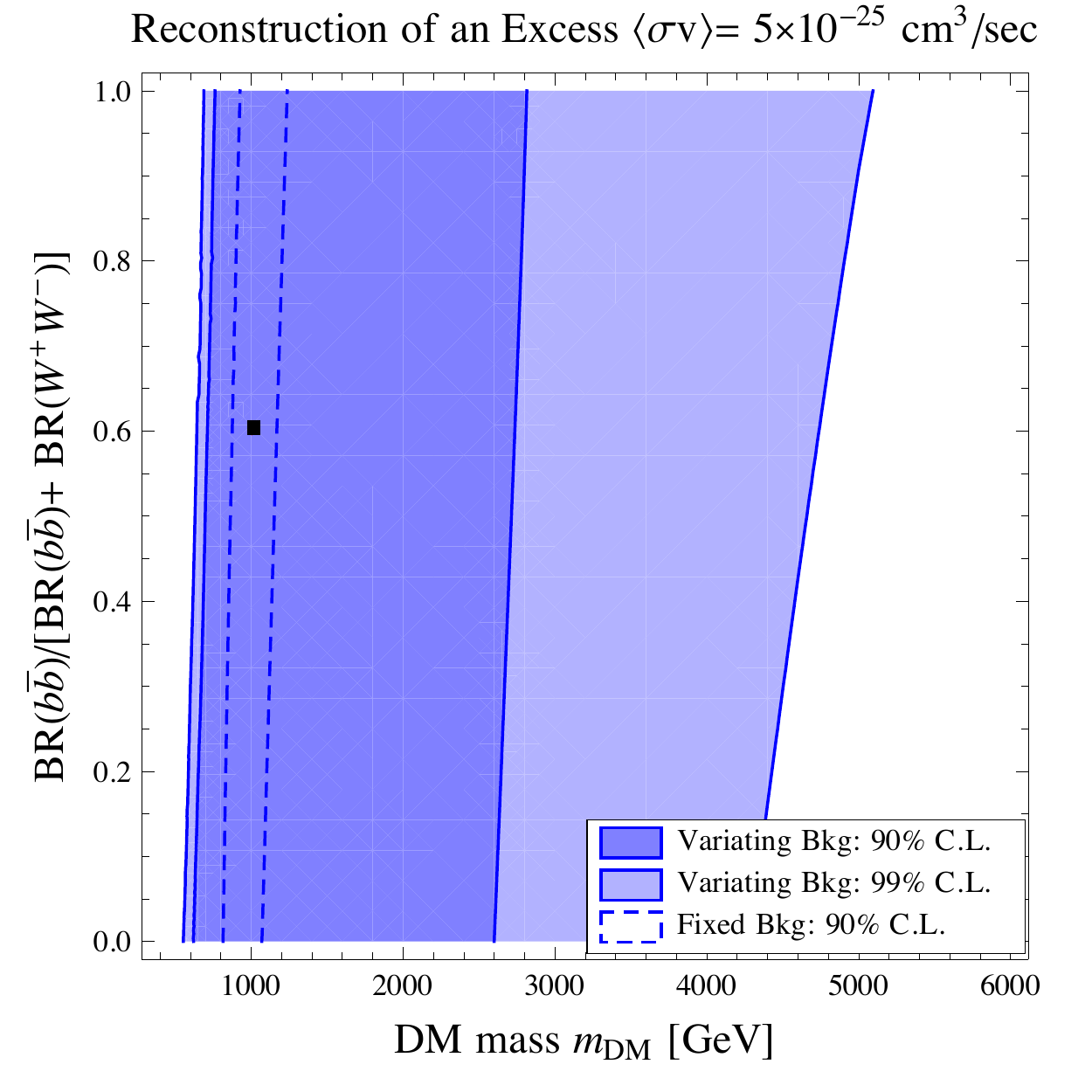}
\caption{\small \em\label{fig:excess_msr} {\bf Reconstruction capabilities of DM properties by \AMS.} {\em Left Panels:} Reconstruction of the mass and the cross-section assuming an annihilation into $b\bar{b}$. The true point is indicated by a symbol; the contours delimit the regions that \AMS\ would identify, at the indicated C.L. and given the indicated assumptions on the background. {\em Middle Panel:} Reconstruction of the mass and the cross-section without knowing a priori the annihilation channel. The true point is indicated by the symbol; the other points and the contours indicate the best fits and the 90\% C.L. regions that \AMS\ would identify for other annihilation channels. {\em Right Panels:} Reconstruction of the mass and the branching ratio for a fixed cross-section and annihilation into $b\bar{b}$ and $W^+W^-$. The symbols represent the model used to produce the mock data.}
\end{center}
\end{figure}

We need to assume values of the DM mass and annihilation cross-section which are still allowed by the \PAMELA\ data, but which would be probed by \AMS. We therefore select three benchmark models:
\begin{itemize}
\item[\footnotesize$\blacktriangle$] $m_{\rm DM}= 85 \GeV \qquad  \langle \sigma v \rangle= 2.7 \times 10^{-26} \, {\rm cm}^3\, {\rm s}^{-1}$,
\item[\footnotesize $\bigstar$] $m_{\rm DM}= 300 \GeV \qquad  \langle \sigma v \rangle= 1.5 \times 10^{-25} \, {\rm cm}^3\, {\rm s}^{-1}$,
\item[\scriptsize$\blacksquare$] $m_{\rm DM}= 1 \TeV \qquad  \langle \sigma v \rangle= 5 \times 10^{-25} \, {\rm cm}^3\, {\rm s}^{-1}$,
\end{itemize}
which are denoted by the corresponding symbols in Fig.s~\ref{fig:constraints_ann}, \ref{fig:AMS_ann_decay} and \ref{fig:excess_msr}. Each of these models represents a different region of interest in the ($m_{\rm DM}$,$\langle \sigma v\rangle$)-plane: 
\begin{itemize}
\item[\footnotesize$\blacktriangle$] The cross-section corresponds to the thermal annihilation cross-section, favored by cosmological observations. Because of the relatively small mass, the DM signal affects mainly energies below 10 GeV (which are not considered for the analysis).
\item[\footnotesize $\bigstar$] The $\bar{p}$ signal sits squarely in the energy range probed by \AMS\ . For the given mass, the cross-section is chosen at the limit of the exclusion contour of PAMELA. The analysis of this model should be the most straightforward.
\item[\scriptsize$\blacksquare$] The DM signal starts to have an important contribution for energies around 100 GeV, where larger uncertainties are present. The lack of data for high energies should pose problem for the reconstruction of this model.
\end{itemize}

\noindent We assume an Einasto halo profile and we fix `MED' propagation parameters. We sum the $\bar p$ flux resulting from these DM models to the `fixed background' and we generate the corresponding mock data, plotted in Fig.~\ref{fig:flux_data}, right. 
We thus determine {\em a posteriori} the regions of the parameter space which would be identified, at a given C.L., by a blind analysis of such data.

\medskip

\begin{table}[p]
\small
\center

\begin{tabular}{|l|c|c|c|}
\hline
Fixed annihilation channel& mass $m_{\rm DM}$& cross-section $\langle \sigma v \rangle$&$\Delta \chi^2$ with respect\\
(true signal: {\boldmath $100\% \ b \bar b$}) & [GeV]& [cm$^3$s$^{-1}$]&to a pure background\\
\hline
$\chi \chi \rightarrow b\bar{b}$& 84.9& $2.7\cdot10^{-26} $ & -10.5\\ 
$\chi \chi \rightarrow W^+ W^-$&84.8 &$6.6\cdot 10^{-26}$&-10.3\\ 
$\chi \chi \rightarrow \gamma \gamma$&57.4&$1.0\cdot 10^{-24}$&-5.7\\ 
\hline
\hline
Fixed cross-section $\langle\sigma v\rangle$ [cm$^3$s$^{-1}$]& mass $m_{\rm DM}$& relative & $\Delta \chi^2$ with respect\\
(true signal: {\boldmath $70\% \, b \bar b + 30\%\, W^+W^-$})& [GeV] & branching ratio &to a pure background\\
\hline
$2.7\times 10^{-26}$& 84.9 &0.7& $ -5.4 $\\
\hline
\end{tabular}
\caption{\small\em\label{tab:chi2_85GeV}Best fit reconstructions of a signal from a 85 GeV DM candidate with $\langle\sigma v\rangle=2.7 \times 10^{-26}\ cm^3\,s^{-1}$ (model {\footnotesize$\blacktriangle$}). See text for details.} 

\vspace{0.8cm}

\begin{tabular}{|l|c|c|c|}
\hline
Fixed annihilation channel& mass $m_{\rm DM}$& cross-section $\langle \sigma v \rangle$&$\Delta \chi^2$ with respect\\
(true signal: {\boldmath $100\% \ b \bar b$}) & [GeV]& [cm$^3$s$^{-1}$]&to a pure background\\
\hline
$\chi \chi \rightarrow b\bar{b}$&300 &$1.5 \cdot 10^{-25}$ & -21.0\\ 
$\chi \chi \rightarrow W^+ W^-$&240 &$1.9 \cdot 10^{-25}$&-19.7\\ 
$\chi \chi \rightarrow \gamma \gamma$&169 &$ 4.8 \cdot  10^{-24}$&-9.8\\ 
$\chi \chi \rightarrow \mu^+ \mu^-$&447&$7.6 \cdot  10^{-23}$&-19.2\\ 
\hline
\hline
Fixed cross-section $\langle\sigma v\rangle$ [cm$^3$s$^{-1}$]& mass $m_{\rm DM}$& relative & $\Delta \chi^2$ with respect\\
(true signal: {\boldmath $40\% \, b \bar b + 60\%\, W^+W^-$})& [GeV] & branching ratio &to a pure background\\
\hline
$1.5\times 10^{-25}$& 300& 0.4& $-13.0$\\
\hline
\end{tabular}
\caption{\small\em\label{tab:chi2_300GeV}Best fit reconstructions of a signal from a 300 GeV DM candidate with $\langle\sigma v\rangle=1.5 \times 10^{-25}\ cm^3\,s^{-1}$ (model {\footnotesize$\bigstar$}). See text for details.} 

\vspace{0.8cm}

\begin{tabular}{|l|c|c|c|}
\hline
Fixed annihilation channel& mass $m_{\rm DM}$& cross-section $\langle \sigma v \rangle$&$\Delta \chi^2$ with respect\\
(true signal: {\boldmath $100\% \ b \bar b$}) & [GeV]& [cm$^3$s$^{-1}$]&to a pure background\\
\hline
$\chi \chi \rightarrow b\bar{b}$& 999 &$ 5 \cdot 10^{-25}$ & -15.7\\ 
$\chi \chi \rightarrow W^+ W^-$&886 &$ 5.8 \cdot 10^{-24}$&-15.1\\ 
$\chi \chi \rightarrow \gamma \gamma$&765&$ 2.5 \cdot 10^{-22}$&-14.5\\ 
$\chi \chi \rightarrow \mu^+ \mu^-$&1711&$ 1.1 \cdot 10^{-22}$&-14.8\\ 
\hline
\hline
Fixed cross-section $\langle\sigma v\rangle$ [cm$^3$s$^{-1}$]& mass $m_{\rm DM}$ & relative & $\Delta \chi^2$ with respect\\
(true signal: {\boldmath $60\% \, b \bar b + 40\%\, W^+W^-$}) & [GeV] & branching ratio & to a pure background\\
\hline
$5\times 10^{-25}$& 999& 0.6& $-15.3$\\
\hline
\end{tabular}
\caption{\small\em\label{tab:chi2_1TeV}Best fit reconstructions of a signal from a 1 TeV DM candidate with $\langle\sigma v\rangle=5 \times 10^{-25}\ cm^3\,s^{-1}$ (model {\scriptsize$\blacksquare$}). See text for details.}
\end{table}

\subsubsection*{Mass and cross-section for $b\bar{b}$ channel}
First we assume that the annihilation channel is fixed and known: $\chi \bar \chi \to b \bar b$ with 100\% Branching Ratio (BR). 
In the left panels of Fig.~\ref{fig:excess_msr} we show the regions identified at 90\% and 99\% C.L. (corresponding to $\Delta \chi^2 = 4.61, 9.21$ for 2 d.o.f.), with the ellipse corresponding to the `fixed background' hypothesis and the extended shaded areas corresponding to the implementation of the `variating background' procedure. 
\begin{itemize}
\item[\footnotesize $\bigstar$] We focus first on the results for this model. We see that $m_{\rm DM}$ is determined fairly well, within 50\% of the true value or so. For a fixed cross section, values of the mass which are smaller than the true value are more quickly disfavored than values which are larger than the true one. This is consistent with expectations since, for small masses, the signal will have its maximum at small energies, which are measured more precisely. 
\\ We also notice that the ellipse corresponding to the `fixed background' analysis displays the degeneracy between $m_{\rm DM}^{-2}$ and $\langle \sigma v \rangle$, since the product of these two quantities appears in the determination of the DM flux. On the other hand, the `variating background' analysis destroys this simple analytical dependence.
\item[\footnotesize$\blacktriangle$] Here, we display only the 90\% C.L. contours: the 99\% ones would artificially extend to very low DM masses, as a consequence of the 10 GeV cut imposed by solar modulation. From the point of view of the $\chi^2$, the points in the 99\% contour would favor a pure background. The shape is unusual probably because of the cut at 10 GeV.
\item[\scriptsize$\blacksquare$] The 99\% C.L. region extends to very large masses as the signal moves into the region in which the measurements are least precise. In addition, as the signal affects large kinetic energies, any contribution with a mass $>$ 1 TeV can fit the the data at 99\% C.L. provided the cross-section is large enough. 
\end{itemize}
In all cases, the annihilating cross section is determined within an order of magnitude at best.

\subsubsection*{Mass and the cross-section for different annihilation channels}
Next, we use still a single annihilation channel ($\chi \bar \chi \to b \bar b$ with 100\% BR) to produce the mock data and we investigate how the same signal could be interpreted in terms of a different channel. The results are shown in the central panels of Fig.~\ref{fig:excess_msr}, where the contours enclose the 90\% C.L. regions for the indicated channels. Not surprisingly, interpreting the data in terms of annihilation into the $W^+W^-$ channel identifies a region which is very similar (a part for the kinematical cut in the {\footnotesize$\blacktriangle$} case) to the one corresponding to the `true' $b \bar b$ annihilation, since the $\bar p$ spectra and the yields of the two channels are very similar. On the other hand, interpreting the data in terms of annihilation into $\gamma\gamma$ ($\mu^+\mu^-$) would result in reconstructing a DM mass smaller (larger) than the true one and a cross section much larger than the true one. 
In Tables~\ref{tab:chi2_300GeV} and~\ref{tab:chi2_1TeV} (upper parts) we report the best fit values of DM mass and annihilation cross section (identified by a colored point in Fig.~\ref{fig:excess_msr}) for each channel. We also report, in the last column of the Tables, the $\Delta \chi^2$ of each one of these points with respect to a pure background hypothesis.\,\footnote{More precisely, what we do is the following. We generate the mock data using a `fixed background + signal' flux. We then try and interpret them in terms of a `variating background' with free $A$ and $p$: the best of such variating backgrounds has a $\chi^2$ of $\chi^2_0$. We then add back the signal, assuming different annihilation channels, and we see how much the $\chi^2$ improves with respect to $\chi^2_0$. These improvements are the numbers we quote in the last column of the tables.} 
We see that, while the $b \bar b$ channel does still obtain the overall best fit, all the other channels also fit the data similarly well, so that it is not practically possible to reconstruct the `true' configuration. In addition, we remind that a channel like $t \bar t$ (when kinematically open) is practically indistinguishable from $b \bar b$ and therefore constitutes an additional degeneracy.

The $\Delta \chi^2$ numbers allow to draw another conclusion: all channels for the {\footnotesize$\bigstar$} and {\scriptsize$\blacksquare$} models (with the possible exception of the borderline case of the $\gamma\gamma$ channel in the {\footnotesize$\blacktriangle$} model) improve sufficiently the $\chi^2$ that an excess can quite clearly be identified. This might be surprising if, for instance, judging by eye from the right panel of fig.~\ref{fig:flux_data}. However, as long as a power-law assumption (with the assumed uncertainty band) for the background is justified, this statistical conclusion is sound. If one allows for the possibility that the background shows features mimicking the hump produced by DM (or that it might consist of a superposition of power-laws, or that the uncertainty band in much larger...), then of course these conclusions evaporate, as warned in Sec.~\ref{secondaries}.
On the other hand, for the {\footnotesize$\blacktriangle$} model the improvement of $\chi^2$ over the pure background becomes quite thin: an actual reconstruction will be very challenging.

\subsubsection*{Branching Ratio}
Finally, we produce mock data with a mixed annihilation into $b \bar b$ and into $W^+W^-$ and we investigate whether it is possible to reconstruct the Branching Ratio. The results in the right panels of fig.~\ref{fig:excess_msr} show that it is not possible. While the determined best fit falls on top or very close to the true value (see the last lines of the tables~\ref{tab:chi2_85GeV}, \ref{tab:chi2_300GeV}, \ref{tab:chi2_1TeV}), the 90\% and 99\% C.L. regions span the entire range of BR. 
This is due to the similarity of the $b\bar b$ and $W^+W^-$ spectra and it reinforces the conclusions of the analyses above. In particular, these conclusions highlight that it will be important to carry on a {\em multi-messenger} analysis (i.e. including signals in $\gamma$ rays, neutrinos etc) in order to break the degeneracies and obtain a convincing reconstruction.

\section{Conclusions}
\label{conclusions}

In this paper we have explored the potential of the current and future measurements of the flux of CR antiprotons to constrain Dark Matter. We have analyzed, in a model independent way, several annihilation and decay channels, a large range of masses and assessed the impact of astrophysical uncertainties. Our main results can be schematized as follows.
\begin{itemize}
\item[$-$] The current $\bar p$ constraints using \PAMELA\ measurements are very strong both for annihilating (see fig.~\ref{fig:constraints_ann} left) and decaying (see fig.~\ref{fig:constraints_decay} left) Dark Matter, for the hadronic channels. Adopting fiducial choices for the halo profile and propagation parametres (`Einasto, MED'), they are as strong as (or even slightly stronger than) the most stringent gamma ray constraints from the \Fermi\ satellite. 
\item[$-$] The astrophysical uncertainty band associated to different choices for the halo profile and the propagation parameters, however, spans between one and two orders of magnitude (see fig.~\ref{fig:constraints_ann} right and fig.~\ref{fig:constraints_decay} right). 
\item[$-$] The upcoming \AMS\ measurements (assuming 1 year of accumulated data and a realistic acceptance for the experiment) have the power to improve the constraints, assuming that no signal is seen, by slightly less than one order of magnitude (see fig.~\ref{fig:AMS_ann_decay}). In particular, \AMS\ will probe the thermal value of the annihilation cross section in the range $m_{\rm DM} \simeq 30 - 200$ GeV for the $b \bar b$ (or $t \bar t$) channel.
\item[$-$] If a convincing excess is measured by \AMS\ and it is attributed to annihilating Dark Matter, the data will allow to somewhat constrain the underlying DM properties, but a full reconstruction will be quite challenging (see fig.~\ref{fig:excess_msr}). The main reason is that the $\bar p$ spectra from different channels are quite similar one to another (as a result of the hadronization process smoothing out differences) and not very different from the shape of the pure background, so that degeneracies are impossible to break.
\begin{itemize}
\item[$-$] Restricting to the `traditional' hadronic annihilation channels ($b\bar b$, $t\bar t$, $W^+W^-$, $ZZ$), in the most favorable case in which $m_{\rm DM} \sim$ few hundreds GeV (and therefore the signal falls in the energy range best measured by \AMS) it will be possible to reconstruct the DM mass within 50\%. For smaller values of $m_{\rm DM}$ (below $\sim 100$ GeV) or larger ones (above $\sim 1$ TeV) the reconstruction capability worsens somewhat.
\item[$-$] Still restricting to the hadronic channels, the determination of the annihilation cross section will be possible within an order of magnitude independently of the mass range.
\item[$-$] If other annihilation channels are allowed, the data will not allow to discriminate among them and the determination of the annihilation cross section becomes impossible.
\item[$-$] If several of the hadronic channels are open at the same time, it will not be possible to determine the relative branching ratios.
\end{itemize} 
\end{itemize}

\noindent In summary, we find that antiprotons are a very relevant tool to constrain Dark Matter annihilation and decay, on a par with gamma rays for the hadronic channels. Current \PAMELA\ data and especially upcoming \AMS\ data allow to probe large regions of the parameter space. On the other hand, using $\bar p$ to reconstruct the DM properties in case of positive detection will be very challenging, except for favorable scenarios.

\paragraph{Acknowledgements}
We thank Massimo Gervasi and Sonia Natale for useful clarifications concerning \AMS\ and Ilias Cholis, Carmelo Evoli and Dario Grasso for communications on their previous work.
We also thank Pasquale D. Serpico and Marco Taoso for useful discussions and for reading the manuscript.
This work is supported by the European Research Council (ERC) under the EU Seventh Framework Programme (FP7/2007-2013) / ERC Starting Grant (agreement n. 278234 - `NewDark' project).
The work of MC is also supported in part by the French national research agency ANR under contract ANR 2010 BLANC 041301 and by the EU ITN network UNILHC. 


\bigskip
\appendix

\footnotesize
\begin{multicols}{2}
  
\end{multicols}

\end{document}